\documentclass[aps,prd,preprint,showpacs,eqsecnum,nofootinbib]{revtex4}
\usepackage{graphicx,epsf}
\newif\ifdraft
\drafttrue
\newif\ifpreprint
\preprinttrue

\def\fig#1{fig.~{\ref{#1}}}
\def\Fig#1{Fig.~{\ref{#1}}}

\def\sect#1{section~{\ref{#1}}}

\def\Tab#1{Table~{\ref{#1}}}

\def\spa#1.#2{\left\langle#1\,#2\right\rangle}
\def\spb#1.#2{\left[#1\,#2\right]}
\def\spash#1.#2{\spa{\smash{#1}}.{\smash{#2}}}
\def\spbsh#1.#2{\spb{\smash{#1}}.{\smash{#2}}}
\def\lor#1.#2{\left(#1\,#2\right)}
\def\sand#1.#2.#3{%
\left\langle\smash{#1^{-}}{\vphantom1}\right|{#2}%
\left|\smash{#3^{-}}{\vphantom1}\right\rangle}
\def\sandpp#1.#2.#3{%
\left\langle\smash{#1^{+}}{\vphantom1}\right|{#2}%
\left|\smash{#3^{+}}{\vphantom1}\right\rangle}
\def\sandpm#1.#2.#3{%
\left\langle\smash{#1^{+}}\vphantom1\right|{#2}%
\left|\smash{#3^{-}}\vphantom1\right\rangle}
\def\sandmp#1.#2.#3{%
\left\langle\smash{#1^{-}}\vphantom1\right|{#2}%
\left|\smash{#3^{+}}{\vphantom1}\right\rangle}
\def\sandmppm#1.#2.#3{%
\left\langle\smash{#1^{\mp}}\vphantom1\right|{#2}%
\left|\smash{#3^{\pm}}{\vphantom1}\right\rangle}
\def\sandnn#1.#2.#3{%
\left\langle\smash{#1}\vphantom1\right|{#2}%
\left|\smash{#3}{\vphantom1}\right\rangle}
\def\sandmn#1.#2.#3{%
\left\langle\smash{#1^{-}}\vphantom1\right|{#2}%
\left|\smash{#3}{\vphantom1}\right\rangle}
\def\sandnm#1.#2.#3{%
\left\langle\smash{#1}\vphantom1\right|{#2}%
\left|\smash{#3^{-}}{\vphantom1}\right\rangle}

\def\tree{{\rm tree}}

\def\e{\epsilon}
\def\eps{\epsilon}
\def\ep{\epsilon}

\def\nn{\nonumber}
\def\Res{\mathop{\rm Res}}
\def\BlackHat{{\tt BlackHat}}

\def\eqn#1{eq.~(\ref{#1})}
\def\Eqn#1{Equation~(\ref{#1})}
\def\eqns#1#2{eqs.~(\ref{#1}) and~(\ref{#2})}

\def\be{\begin{equation}}
\def\ee{\end{equation}}
\def\bea{\begin{eqnarray}}
\def\eea{\end{eqnarray}}
\def\ba{\begin{eqnarray}}
\def\ea{\end{eqnarray}}

\def\tree{{\rm tree}}
\def\oneloop{{\rm 1\hbox{-}loop}}

\newbox\charbox
\newbox\slabox
\def\s#1{{      % Feynman slash
        \setbox\charbox=\hbox{$#1$}
        \setbox\slabox=\hbox{$/$}
        \dimen\charbox=\ht\slabox
        \advance\dimen\charbox by -\dp\slabox
        \advance\dimen\charbox by -\ht\charbox
        \advance\dimen\charbox by \dp\charbox
        \divide\dimen\charbox by 2
        \raise-\dimen\charbox\hbox to \wd\charbox{\hss/\hss}
        \llap{$#1$} }}

\newcommand{\Bmp}[1]{\langle #1\rangle}

\newcommand{\Kf}[1]{\tilde K_{#1}}
\newcommand{\Kfmu}[1]{\tilde K^{\mu}_{#1}}

\newcommand{\Kfm}[1]{\tilde K^{-}_{#1}}

\begin{document}

%\iffalse
\ifpreprint
UCLA/08/TEP/10
\hfill $\null\hskip 1.0 cm \null$ MIT-CTP-3937
\hfill $\null\hskip 1.0 cm \null$ NSF-KITP-08-48\\
$\null\hskip .4 cm$ Saclay-IPhT--T08/054
\hfill $\null\hskip 1.0 cm \null$ SLAC--PUB--13161
\fi
%\fi

\title{An Automated Implementation of On-Shell Methods for
          One-Loop Amplitudes}

\author{C.~F.~Berger${}^{a,b,c}$, Z.~Bern${}^d$, L.~J.~Dixon${}^c$,
F.~Febres Cordero${}^d$, D.~Forde${}^{c,d}$, H. Ita${}^d$, 
D.~A.~Kosower${}^{e}$ and D.~Ma\^{\i}tre${}^c$}%
\affiliation{\centerline{${}^a${Center for Theoretical
Physics, Massachusetts Institute of Technology,
      Cambridge, MA 02139, USA}} \\
\centerline{%
${}^b${Kavli Institute for Theoretical Physics, University of California,
Santa Barbara, CA 93106, USA}} \\
\centerline{${}^c$Stanford Linear Accelerator Center, Stanford University,
             Stanford, CA 94309, USA} \\
\centerline{${}^d$Department of Physics and Astronomy, UCLA, Los Angeles, CA
90095-1547, USA} \\
\centerline{${}^e$Institut de Physique Th\'eorique, CEA--Saclay,
          F--91191 Gif-sur-Yvette cedex, France}\\
}

\begin{abstract}
  We present the first results from \BlackHat{}, an automated C++
  program for calculating one-loop amplitudes. The program implements
  the unitarity method and on-shell recursion to construct
  amplitudes.  As input to the calculation, it uses
  compact analytic formul\ae{} for tree amplitudes for 
  four-dimensional helicity states.  The program
  performs all related computations numerically.  We make
  use of recently developed on-shell methods for evaluating coefficients
  of loop integrals, introducing a discrete Fourier projection as a means 
  of improving efficiency and numerical stability.  
  We illustrate the numerical stability of our approach by
  computing and analyzing six-, seven- and eight-gluon amplitudes in QCD
  and comparing against previously-obtained analytic results.
\end{abstract}


\maketitle

\section{Introduction}
\label{IntroSection}

The Large Hadron Collider (LHC) will soon begin exploration of the
electroweak symmetry breaking scale. It is widely anticipated that physics
beyond the Standard Model will emerge at this scale, leading to a
breakthrough in our understanding of TeV-scale physics. A key ingredient
in this quest is the precise understanding of the expected Standard Model
backgrounds to new physics from both electroweak and QCD processes.  In
the absence of such an understanding, new physics signals may remain
hidden, or backgrounds may be falsely identified as exciting new physics
signals.

Quantitatively reliable QCD predictions require next-to-leading order
(NLO) calculations~\cite{NLMLesHouches}.  For a few benchmark
processes, such as the rapidity distribution of electroweak vector
bosons~\cite{NNLOVrap}, the transverse-momentum distribution of the
$Z$ boson at moderate $p_{\rm T}$, and the total cross sections for
production of top quark pairs and of Higgs bosons~\cite{HNNLO}, the
higher precision of next-to-next-to-leading order (NNLO) results may
be required~\cite{NLMLesHouches}.  For most other processes, NLO
precision should suffice.  However, there are many relevant processes
that need to be computed, particularly those with high final-state
multiplicity.  Such processes are backgrounds to the production of new
particles that have multi-body decays.  To date, no complete NLO QCD
calculation involving four or more final-state objects (particles or
jets) is available.  (In electroweak theory, however, $e^+e^-
\rightarrow 4$ fermions has been evaluated~\cite{ElectroWeak4f} using
the integral reduction scheme of Denner and Dittmaier~\cite{Denner}.)
NLO corrections require as ingredients both real-radiative corrections
and virtual corrections to basic amplitudes.  The structure of the
real-radiative corrections --- isolation of infrared singularities and
their systematic cancellation against virtual-correction singularities
--- is well understood, and there are general methods for organizing
them~\cite{GG,GGK,CS}.  Indeed, the most popular of these methods, the
Catani-Seymour dipole subtraction method~\cite{CS}, has now been
implemented in an automatic fashion~\cite{GK}.  The infrared
divergences of virtual corrections, needed to cancel the divergences
from integrating real radiation over phase space, are also understood
in general~\cite{GG,KST}.  The main bottleneck to NLO computations of
processes with four or more final-state objects has been the
evaluation of the remaining ingredients, the infrared-finite parts of
the one-loop virtual corrections.

As the number of external particles increases, the computational
difficulty of loop-amplitude calculations using traditional Feynman
diagrams grows rapidly.  Technologies that have proven useful at tree
level, such as the spinor-helicity formalism~\cite{SpinorHelicity}, 
do not suffice to tame these
difficulties.  In the past few years, several classes of new methods
have been proposed to cope with this rapid
growth~\cite{OtherApproaches,AguilaPittau,Binoth,Denner,EGZ,XYZ},
including on-shell methods~\cite{UnitarityMethod,Zqqgg,
DDimUnitarity,TwoLoopAllPlus,BCFUnitarity,OtherUnitarity,BCFRecursion,BCFW,
DoublePole,Bootstrap,Genhel,OtherBootstrap,BCFCutConstructible,BMST,
OPP,BFMassive,OnShellReview,Forde,EGK,CutTools,GKM,OPP2,MOPP} 
which are based on the analytic properties of unitarity and
factorization that any amplitude must
satisfy~\cite{OldUnitarity,GeneralizedUnitarityOld}.  These methods
are efficient, and display very mild growth in required computer time
with increasing number of external particles, compared to a
traditional Feynman-diagrammatic approach.  The improved efficiency
emerges from effectively reducing loop calculations to tree-like
calculations.  Efficient algorithms can then be employed for the
tree-amplitude ingredients.

One of the principal on-shell technologies is the unitarity method,
originally developed in calculations of supersymmetric amplitudes with more
than four%
\footnote{The earlier dispersion relation approach~\cite{OldUnitarity} had
not been used to construct amplitudes with more than two kinematic
invariants.}  external particles~\cite{UnitarityMethod,Fusing}.  An early
version combining unitarity with factorization properties was used to
compute the one-loop amplitudes for $e^+e^- \rightarrow Z \rightarrow 4$
partons and (by crossing) for amplitudes entering $pp \rightarrow W,Z$ + 2
jets~\cite{Zqqgg}.  (The latter have been incorporated into the NLO
program {\tt MCFM}~\cite{MCFM}.)  This calculation introduced the concept of
generalized unitarity~\cite{GeneralizedUnitarityOld} as an efficient means
for performing loop computations.  It improves upon basic unitarity because
it isolates small sets of terms, and hence makes use of simpler on-shell
amplitudes as basic building blocks.  On-shell methods have already led to
a host of new results at one loop, including the computation of non-trivial
amplitudes in QCD with an arbitrary number of external
legs~\cite{DoublePole,Bootstrap,OneLoopMHV,Genhel,OtherBootstrap}.  This
computation goes well beyond the scope of traditional diagrammatic
computations, and provides a clear demonstration of the power of the
methods.  The reader may find recent reviews and further references in
refs.~\cite{OnShellReview,NLMLesHouches}.

The next challenge is to move beyond analytic calculations of specific
processes or classes of processes to produce a complete, numerically
stable, efficient computer code based on these new developments.  Here we
report on an automated computer program --- \BlackHat{} --- based on
on-shell methods, with the stability and efficiency required to compute
experimentally-relevant cross sections.  Other researchers are
constructing numerical programs~\cite{EGK,CutTools,GKM,OPP2,MOPP}
based on related methods~\cite{OPP,GKM,MOPP}.

On-shell methods rely on the unitarity of the theory~\cite{OldUnitarity}
and on its factorization properties, which together require that the poles
and branch cuts of amplitudes correspond to the physical propagation of
particles.  In general, any one-loop amplitude computed in a quantum field
theory contains terms with branch cuts, and also purely rational terms,
that is, terms that have no branch cuts and are rational functions of the
external momentum invariants (or more precisely of spinor products).  The
cut-containing pieces can be determined from unitarity cuts, in which the
intermediate states may be treated
four-dimensionally~\cite{UnitarityMethod,Fusing}.  Only products of
tree-level, four-dimensional helicity amplitudes are needed for this step.
The rational terms have their origin in the difference between $D=4-2\e$
and four dimensions when using dimensional regularization. They can be
obtained%
\footnote{This fact is closely connected to van Neerven's important
observation that dispersion relations for Feynman integrals converge in
dimensional regularization~\cite{VanNeervenUnitarity}.}  within the
unitarity method by keeping the full $D$-dimensional dependence of the
tree amplitudes~\cite{DDimUnitarity,TwoLoopAllPlus,BMST,OPP,
BFMassive,GKM,OPP2}.  Alternatively, to obtain the rational terms, one can use
on-shell recursion~\cite{BCFRecursion,BCFW} to construct the rational
remainder from the loop amplitudes' factorization
poles~\cite{Bootstrap,OneLoopMHV,OtherBootstrap}.
We will follow the latter route in this paper.

A generic one-loop amplitude can be expressed in terms of a set of
scalar master integrals multiplied by various rational coefficients,
along with the additional purely rational
terms~\cite{IntegralReductions,BDKIntegrals,FJT,BGH,DuplancicNizic}. 
The relevant master integrals depend on the masses of the physical states
that appear, but otherwise require no process-specific computation.
At one loop, they consist of box, triangle, bubble and (for massive
particles) tadpole integrals.  The required integrals are known
analytically~\cite{IntegralsExplicit,BDKPentagon}.

Our task is therefore to determine the coefficients in front of these
integrals for each process and helicity configuration. We do so using
generalized cuts~\cite{Zqqgg,TwoLoopAllPlus,OtherGeneralized,BCFUnitarity}. 
Britto, Cachazo and Feng (BCF) observed~\cite{BCFUnitarity} 
that with {\it complex\/} momenta
one can use quadruple cuts to solve for {\it all\/} box coefficients,
because massless three-point amplitudes isolated by cuts do not vanish
as they would for real massless momenta.  Moreover, the solution is purely
algebraic, because the loop momentum of the four-dimensional integral
is completely frozen by the four cut conditions, and a given
quadruple cut isolates a unique box coefficient. This provides an
extremely simple method for computing box-integral coefficients.
Continuing along these lines, Britto, Buchbinder, Cachazo, Feng, and
Mastrolia have developed efficient analytic
techniques~\cite{BCFCutConstructible} for evaluating generic one-loop
unitarity cuts to compute triangle and bubble coefficients.  They use
spinor variables and compute integral coefficients
via residue extraction.

For the purposes of constructing a numerical code, we use a somewhat
different approach.  For triangle integrals, we can impose at most
three cut conditions.  This leaves a one-parameter family of
solutions.  These conditions no longer isolate the triangle integral
uniquely, as a number of box integrals will share the same triple cut.
Similar considerations apply to the ordinary two-particle cuts needed
to obtain bubble coefficients.  As discussed by Ossola, Papadopoulos
and Pittau (OPP)~\cite{OPP}, one can construct a general parametric
form for the integrand.  This form can be understood as a
decomposition of the loop momentum in terms of components in the
hyperplane of external momenta and components perpendicular to this
hyperplane~\cite{EGK}.  Coefficients of the various master integrals
can be extracted by comparing the expressions obtained from Feynman
graphs with the general parametric form, using values of the loop
momentum in which different combinations of propagators go on shell.
For the quadruple cut, this leads to a computation identical to the
method of ref.~\cite{BCFUnitarity} once one further replaces sums of
Feynman diagrams by tree amplitudes.  OPP solve the problems of box
contributions to triangle coefficients, and of box and triangle contributions
to bubble coefficients, iteratively by subtracting off
previously-determined contributions and solving a particular system of
equations numerically. In the OPP approach, the rational terms can be
determined by keeping the full $D$-dimensional dependence in all
terms~\cite{OPP,GKM,OPP2}.

Forde's alternative approach makes use of a complex-valued parametrization
of the loop momenta~\cite{Forde} (similar to the one used in
refs.~\cite{AguilaPittau,OPP}) and exploits the different functional
dependence on the complex parameters to separate integral
contributions to a given triple or ordinary cut. We develop this
method one step further, and introduce a discrete Fourier projection in
these complex parameters, in conjunction with an OPP subtraction of
previously-determined contributions~\cite{OPP}. The projection isolates 
the desired integral coefficients efficiently, while maintaining good
numerical stability in all regions of phase space.  It minimizes the
instabilities that may arise~\cite{EGK,OPP2,MOPP} from solving a system 
of linear equations in regions where the system degenerates.

We compute the rational remainder terms using loop-level on-shell recursion
relations~\cite{Bootstrap,OneLoopMHV,OtherBootstrap}, analogous to
the recursion relations at tree
level~\cite{BCFRecursion,BCFW} developed by
Britto, Cachazo, Feng and Witten (BCFW).  At tree level, gauge-theory
amplitudes can be constructed recursively from lower-point amplitudes,
by applying a complex deformation to the momenta of a pair of external
legs, keeping both legs on shell and preserving momentum conservation.
The proof relies only on the factorization properties of the theory
and on Cauchy's theorem, so the method can be applied to a wide
variety of theories.  At loop level, the construction of an analogous
recursion relation for the rational terms requires addressing a number of
subtleties, including the presence of spurious singularities.  These
issues can and have been addressed for specific infinite series of 
one-loop helicity amplitudes, allowing their recursive 
construction~\cite{Bootstrap,Genhel,OneLoopMHV,OtherBootstrap}.  
In order to more easily automate the method of ref.~\cite{Bootstrap},
we modify how the spurious singularities are treated,
making use of the availability of the integral coefficients 
within the numerical program, in a manner to be described below.

In any numerical method, the finite precision of a computation means
that instabilities can arise, occasionally leading to substantial
errors in evaluating an amplitude at a given point in phase space.  
We introduce simple tests for the stability of the evaluation.  
Principally, we check that the sum of bubble integral coefficients 
agrees with its known value, and we check for the absence of 
spurious singularities in this sum.
A comparison with known analytic answers for a variety of gluon 
amplitudes shows that these two tests suffice to detect almost
all instabilities.  If a test fails,
we consider the point to be unstable.  Various means of
dealing with unstable points have been discussed~\cite{CGM,
GramInterpolate,Denner,FJT,OtherGramMethods,NLMLesHouches}.
We simply re-evaluate the fairly small fraction of
unstable points at higher precision using the {\tt QD}
package~\cite{QD}. Doing so, we still have an average evaluation time
of less than 120 ms for the most complicated of the six-gluon
helicity amplitudes, and subtantially better times for the simpler
ones.  Higher-precision evaluation has also
been used recently in ref.~\cite{CutTools} to handle numerically
unstable points.

Although \BlackHat{} is written in C++, for algorithm development and
prototyping, we found it extremely useful to use symbolic languages
such as Maple~\cite{Maple} and Mathematica~\cite{Mathematica}, and in
particular the Mathematica implementation of the spinor-helicity
formalism provided by the package S@M~\cite{SAM}.  At present
\BlackHat{} computes multi-gluon loop amplitudes.  Once we implement a
wider class of processes in the same framework, we intend to release
the code publicly.

The present paper is organized as follows.  In~\sect{CutsSection}, we
discuss how we compute the coefficients of the various integral
functions, and introduce the discrete Fourier projection.
In~\sect{RationalSection}, we outline the calculation of the
purely-rational terms, describing in particular our treatment of
the spurious singularities.  We also introduce our criteria for ensuring the
numerical stability of the computed amplitude.  We show results for a
number of gluon amplitudes with up to eight external legs
in~\sect{ResultsSection}, and summarize in~\sect{ConclusionsSection}.
We defer a number of technical details to a future paper~\cite{Future}.

%%%%%%%%%%%%%%%%%%%%%%%%%%%%%%%%%%%%%%%%%%%%%%

\section{Integral Coefficients from Four-Dimensional Tree Amplitudes}
\label{CutsSection}

%%% FIGURE %%%%%%%%%%%%%%%%%%%%%%%%%%%%%%%%%%%%%%%%
\begin{figure}
\begin{center}
\includegraphics*[width=0.65\textwidth]{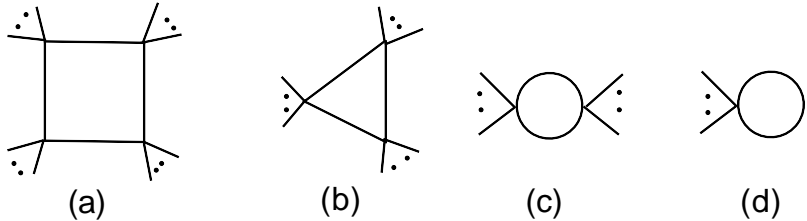}
%\centerline{\epsfxsize 2.5 truein \epsfbox{Cuts.eps}}
 \caption{The basis of scalar integrals: (a) box, (b) triangle, 
(c) bubble, and (d) tadpole.  Each corner can have one or
more external momenta emerging from it.  The tadpole integral (d) vanishes
when all internal propagators are massless.} 
\label{IntegralBasisFigure}
\end{center}
\end{figure}
%%%%%%%%%%%%%%%%%%%%%%%%%%%%%%%%%%%%%%%%%%%%%%%%%%%      

We begin by dividing the dimensionally-regularized amplitude into
cut-containing and rational parts. 
We evaluate the cut parts using the four-dimensional unitarity
method~\cite{UnitarityMethod,OnShellReview}.  To extract the
box-integral coefficients we use the observation of BCF that the
quadruple cuts freeze the loop integration~\cite{BCFUnitarity}.  For
triangle and bubble integrals we use key elements of both the
OPP~\cite{OPP} and Forde~\cite{Forde} approaches.  In addition, we
introduce a discrete Fourier projection for extracting
the integral coefficients.  (Alternative on-shell methods for obtaining 
the integral coefficients have been given in
refs.~\cite{BCFCutConstructible,BFMassive}.)

As the first step, we separate an $n$-point amplitude $A_n$ into a
cut part $C_n$ and a rational remainder $R_n$,
\begin{equation}
A_n = C_n + R_n \,.
\label{CutRational}
\end{equation}
The cut part is given by a linear combination of scalar basis 
integrals~\cite{IntegralsExplicit,IntegralReductions,BDKIntegrals,%
FJT,BGH,DuplancicNizic},
\begin{equation}
C_n = \sum_i d_i I_4^i + \sum_i c_i I_3^i + \sum_i b_i I_2^i 
 + \sum_i a_i I_1^i \,.
\label{IntegralBasis}
\end{equation}
The integrals $I_4^i, I_3^i, I_2^i, I_1^i$ are scalar box, triangle,
bubble and tadpole integrals, illustrated in
\fig{IntegralBasisFigure}. For massless particles circulating in the
loop, the tadpole integrals vanish in dimensional regularization.
The integral coefficients $d_i, c_i, b_i, a_i$ are rational functions of
spinor products and momentum invariants of
the kinematic variables, and are independent of the dimensional 
regularization parameter $\e$.  The
index $i$ runs over all distinct integrals of each type.  The rational
terms $R_n$ are defined by setting all scalar integrals to zero,%
\footnote{All contributions from the scalar integrals in
\eqn{IntegralBasis} are part of $C_n$, including 
all $1/\e^2$ and $1/\e$ pole terms, $\pi^2$ factors, and pieces
arising from the order $\e^0$ term in the scalar bubble integral.}
\begin{equation}
R_n = A_n \Bigr|_{I_m^i \rightarrow 0}.
\label{RatDef}
\end{equation}
Alternatively, the rational terms can be absorbed into the integral
coefficients by keeping their full dependence%
\footnote{The $\e$ dependence leads only to rational
contributions, because it arises from integrals with $(-2\e)$ components 
of loop momenta in the numerator.  Each such integral can be rewritten 
as the product of $\e$ with a higher-dimensional integral, which 
possesses at most a single, ultraviolet pole in $\e$, whose residue 
must be rational~\cite{Mahlon93,DDimUnitarity}.}
on $\e$.

In this paper, we obtain the integral coefficients at $\e=0$ 
by using the unitarity method with four-dimensional loop
momenta. This method allows us to use powerful four-dimensional spinor
techniques~\cite{SpinorHelicity, TreeReview} to greatly simplify the tree
amplitudes that serve as basic building blocks.  We will instead
obtain the rational terms $R_n$ from on-shell
recursion~\cite{BCFRecursion,BCFW,Bootstrap,Genhel}, 
as explained in the subsequent section.

%%% FIGURE %%%%%%%%%%%%%%%%%%%%%%%%%%%%%%%%%%%%%%%%
\begin{figure}
\begin{center}
\includegraphics*[width=0.26\textwidth]{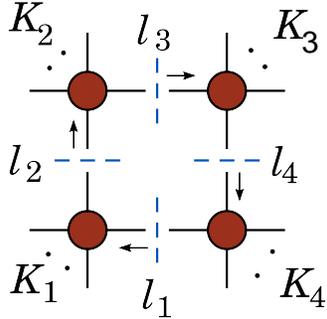}
%\centerline{\epsfxsize 2.5 truein \epsfbox{Cuts.eps}}
 \caption{The quadruple cut used to determine the coefficients of the
box integrals. The loop momenta, flowing clockwise, are constrained to
satisfy on-shell conditions. The blobs at each corner represent tree
amplitudes. The dashed lines indicate the cuts.  The external momenta
are all outgoing.}
\label{QuadCutFigure}
\end{center}
\end{figure}
%%%%%%%%%%%%%%%%%%%%%%%%%%%%%%%%%%%%%%%%%%%%%%%%%%%      

\subsection{Box Coefficients}

Consider first the coefficients of the box integrals.  We obtain them
from the quadruple cut shown in \fig{QuadCutFigure}.  The
cut propagators correspond to the four propagators of the desired box
coefficient.  As observed in ref.~\cite{BCFUnitarity}, if we take the
loop momentum to be four-dimensional, then the four cut conditions,
\begin{equation}
l_i^2 = m_i^2 \,, \hskip 2 cm i = 1,2,3,4,
\label{OnShellEquation}
\end{equation}
match the number of components of the loop momentum,
leading to a discrete sum over two solutions for $l_i$.
The integration is effectively frozen.
The $m_i$ are the masses of the particles in the cut propagators,
which in this paper are taken to vanish.
The coefficient of any box integral is then given in terms of a
product of four tree amplitudes,
\begin{eqnarray}
d_i &=& {1\over 2} \sum_{\sigma = \pm} d_i^\sigma \,,
\label{QuadSum}\\
d_i^\sigma &=& A^\tree_{(1)} A^\tree_{(2)} A^\tree_{(3)} A^\tree_{(4)} 
\Bigr|_{l_i = l_i^{(\sigma)}} \,,
\label{QuadCutSolution}
\end{eqnarray}
where the sum runs over the two solutions to the on-shell conditions,
labeled by ``$+$'' and ``$-$''.  The four tree amplitudes in
\eqn{QuadCutSolution} correspond to the tree amplitudes at the
four corners of the quadruple cut depicted in \fig{QuadCutFigure}.

The generic solution for $l_i^{(\sigma)}$ was found in
ref.~\cite{BCFUnitarity}.   Simpler forms can be found for particular,
but still fairly general, kinematical cases.
In this paper, we focus on the case of massless particles circulating in
the loop.  When in addition at least one external leg, say leg 1,
of the box integral
shown in \fig{QuadCutFigure} is also massless, that is $K_1^2=0$, the
two solutions to the on-shell conditions (\ref{OnShellEquation}) can be
written in a remarkably simple form,
\begin{eqnarray}
&&(l_1^{(\pm)})^\mu =  {\sandmppm1.{\s K_2 \s K_3 \s K_4 \gamma^\mu}.1\over
  2 \sandmppm1.{\s K_2 \s K_4}.1} \,,
\hskip 2cm 
(l_2^{(\pm)})^\mu = 
- {\sandmppm1.{\gamma^\mu \s K_2 \s K_3 \s K_4}.1 \over  
  2 \sandmppm1.{\s K_2 \s K_4}.1} \,,\nn \\
&&(l_3^{(\pm)})^\mu =  
{\sandmppm1.{\s K_2 \gamma^\mu  \s K_3 \s K_4}.1 \over
  2 \sandmppm1.{\s K_2 \s K_4}.1} \,,
\hskip 2cm 
(l_4^{(\pm)})^\mu = 
- {\sandmppm1.{\s K_2 \s K_3 \gamma^\mu  \s K_4}.1\over
  2 \sandmppm1.{\s K_2 \s K_4}.1} \,. \hskip 1 cm 
\label{MasslessSolution}
\end{eqnarray}
As illustrated in \fig{QuadCutFigure}, the $K_i$ are the external
momenta of the corners of the box integral under consideration and
$\langle 1^\mp|$ and $|1^\pm\rangle$ are Weyl spinors corresponding to
the massless momentum $K_1$, in the notation of
refs.~\cite{TreeReview}.  In massless QCD this solution covers all
helicity configurations for amplitudes with 
up to seven external quarks or gluons, and a large fraction of
the box coefficients for more external partons.
(This solution may also be found in Risager's Ph.D.
thesis~\cite{RisagerThesis}.)

The solution~(\ref{MasslessSolution}) has the advantage of making it 
manifest that Gram determinants enter only as {\it square roots}, 
one for each power of loop momenta in the numerator of box integrals.
Indeed, the Gram determinant is given by the product of the 
spinor-product strings in the denominators of \eqn{MasslessSolution},
\begin{eqnarray}
\Delta_4 &=& 
 - 2 \sandmp{1}.{\s K_2 \s K_4}.{1} \sandpm{1}.{\s K_2 \s K_4}.{1} \,,
\end{eqnarray}
where $\Delta_4 = \det(2 K_i \cdot K_j)$, $i,j=1,2,3$, is the box Gram
determinant for $K_1^2=0$.  
This property reduces the severity of numerical round-off 
error due to cancellations between different terms, in the regions of 
phase space where the Gram determinant vanishes.

%%% FIGURE %%%%%%%%%%%%%%%%%%%%%%%%%%%%%%%%%%%%%%%%
\begin{figure}
\begin{center}
\includegraphics*[width=0.7\textwidth]{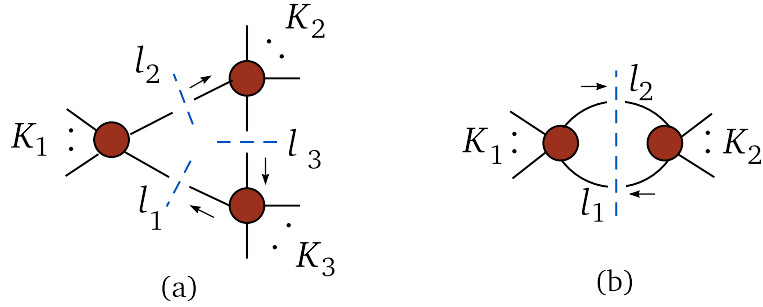}
%\centerline{\epsfxsize 2.5 truein \epsfbox{Cuts.eps}}
\vskip -.3 cm 
 \caption{(a) The triple cut and (b) the ordinary double cut used to
determine the coefficients of the triangle and bubble integrals.  The
loop momenta $l_i$, flowing clockwise, are constrained to satisfy
on-shell conditions. The external momenta are all outgoing. }
\label{TripleDoubleCutFigure}
\end{center}
\end{figure}
%%%%%%%%%%%%%%%%%%%%%%%%%%%%%%%%%%%%%%%%%%%%%%%%%%%      

%%%%%%%%%%%%%%%%%%%%
\subsection{Triangle Coefficients from Discrete Fourier Projection}

To evaluate the coefficients of the triangle and bubble integrals, we
make use of elements from the approaches of both OPP~\cite{OPP} and
Forde~\cite{Forde}.  First consider the triangle integrals. To obtain
the coefficients $c_i$ in \eqn{IntegralBasis} we use the triple cut
depicted in \fig{TripleDoubleCutFigure}(a).  In contrast to the
quadruple cut, the triple cut does not freeze the integral, but leaves
one degree of freedom which we denote by $t$. Moreover, the triple cut
also contains box integral contributions. This makes the extraction of
the triangle coefficients somewhat more intricate than the box
coefficients.

For massless internal particles, the solution of the cut condition $l_i^2 =0$
($i = 1,2,3$) is~\cite{AguilaPittau,OPP,Forde}
\begin{eqnarray}
l^{\mu}_1(t) &=& \Kfmu1 +
\Kfmu3 +
\frac{t}{2}\Bmp{\Kfm1|\gamma^{\mu}|\Kfm3}
+\frac{1}{2t}
\Bmp{\Kfm3|\gamma^{\mu}|\Kfm1}\,,
\label{TripleCutMomentum}
\end{eqnarray}
and, using momentum conservation, $l_2(t)=l_1(t)-K_1$, 
$l_3(t)=l_1(t)+K_3$. 
Here $t$ is a complex parameter corresponding to the one component of the
loop momentum not fixed by the cut condition. Following ref.~\cite{Forde}
we have,
\begin{eqnarray}
\Kfmu1=\gamma\alpha\frac{\gamma K_1^{\mu}+S_1 K^{\mu}_3}{\gamma^2-S_1S_3}\,,
\hskip 2 cm  
\Kfmu3=-\gamma\alpha'\frac{\gamma K_3^{\mu}+S_3 K_1^{\mu}}{\gamma^2-S_1S_3} \,,
\label{eq:def_gamma_pm}
\end{eqnarray} 
with $S_1 = K_1^2$, $S_3 = K_3^2$, and $\Kfmu{1}$ and $\Kfmu{3}$ are both
massless.  (In comparison with ref.~\cite{Forde}, we have rescaled and
relabeled these massless momenta, and here we take all external momenta 
to be outgoing.)
The variables $\alpha$, $\alpha'$ and $\gamma$ are defined as follows,
\begin{eqnarray}
\alpha = \frac{S_3(S_1-\gamma)}{S_1S_3-\gamma^2}\,,\qquad
\alpha' = \frac{S_1(S_3-\gamma)}{S_1S_3-\gamma^2}\,,\qquad
\gamma = \gamma_{\pm} = - K_1\cdot K_3 \pm\sqrt{\Delta}\,,
\end{eqnarray} 
where 
\begin{equation}
\Delta= -\det(K_i\cdot K_j) = (K_1\cdot K_3)^2 - K_1^2K_3^2 \,,
\end{equation}
with $i,j$ running over $1,3$ (or any other pair).  To determine the
coefficients of integrals we must sum over the two solutions
corresponding to $\gamma_+$ and $\gamma_-$.  It turns out that for the
three-external-mass case, these solutions are related by taking
$t\rightarrow 1/t$. In addition, when a corner of the triangle is
massless, simpler forms of the solutions can be obtained.  These
issues will be discussed elsewhere~\cite{Future}.  A
similar solution to \eqn{TripleCutMomentum} has been given in the
massive case~\cite{KilgoreCut}.

OPP~\cite{OPP} showed that after subtracting the known box
contributions from the triple cut integrand, one is left with seven
independent coefficients. One of these seven corresponds to the
coefficient of the scalar triangle we seek, while the remaining six
correspond to terms that integrate to zero. Evaluating the subtracted
triple-cut integrand at seven selected kinematic points leads to a
system of linear equations for these coefficients.  As
discussed in ref.~\cite{EGK}, however, numerical stability issues
can arise from inverting this linear system of equations.  The OPP
approach of solving a system of equations is currently being implemented in
numerical programs, with initial results reported in
refs.~\cite{EGK,CutTools,GKM,OPP2,NLMLesHouches,MOPP}.
In the alternative approach of Forde~\cite{Forde}, the coefficient
is instead extracted from the analytic behavior of the triple cut in the limit
that the complex variable $t$ becomes large.

We choose to use a hybrid of these approaches, subtracting box
contributions from the triple cuts following OPP, but in a way that
makes manifest the analytic properties in the complex variable $t$ following
Forde.  The triple cut is,
\begin{equation}
 C_3(t) \equiv A^\tree_{(1)} A^\tree_{(2)} A^\tree_{(3)} 
\Bigr|_{l_i = l_i(t)} \,.
\label{triplecutC3}
\end{equation}
Each of the box contributions to the triple cut~(\ref{triplecutC3}) 
contains a fourth Feynman propagator, $1/l_i^2(t)$ for some $i\neq1,2,3$.
Hence $C_3(t)$ develops a pole in $t$ whenever the inverse propagator 
vanishes, say
\begin{equation}
l_i^2(t) \sim \xi_i^\sigma (t - t_i^\sigma) \,,  
\hskip .6 cm \hbox{as} \hskip .3 cm 
     t \rightarrow t_i^\sigma\,.
\end{equation}
The pole locations $t_i^\sigma$ and coefficients $\xi_i^\sigma$ 
are determined from the form of $l_i^2(t)$, after inserting
the triple-cut loop momentum parametrization~(\ref{TripleCutMomentum}).
 
\begin{figure}
\begin{center}
\includegraphics*[width=0.25\textwidth]{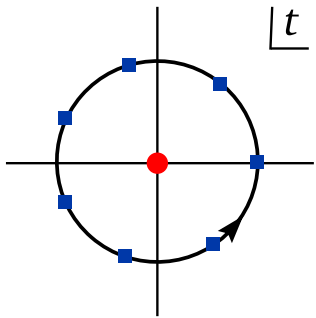}
%\centerline{\epsfxsize 2.5 truein \epsfbox{Contour.eps}}
 \caption{After subtracting the box contributions to the triple cut,
the $t$ plane is free of all singularities except at $t=0$ and 
$t = \infty$.  We can extract the desired triangle coefficient by using 
a discrete Fourier projection, evaluating $T_3(t)$ at points indicated 
by the squares on the circle.}
\label{ContourTriFigure}
\end{center}
\end{figure}

The residues at the poles also involve the coefficients
$d_i^\sigma$ of the $i^{\rm th}$ box integral, evaluated on the 
two solutions $\sigma$ to the quadruple cuts.  The $d_i^\sigma$ can
be computed prior to the triangle calculation, and their contribution
subtracted to form the difference,
\begin{equation}
T_3(t) \equiv C_3(t) 
- \sum_{\sigma=\pm} \sum_i {d_i^\sigma \over \xi_i^\sigma (t - t_i^\sigma)}\,.
\label{T3def}
\end{equation}
\Eqn{T3def} is slightly schematic, omitting a few subtleties that depend
in part on how many of the triangle legs are massive.
For example, in the three-mass case we should either sum over 
$\gamma_+$ and $\gamma_-$, or else make use of the $t\rightarrow 1/t$ 
relation between the 
two triple-cut solutions to eliminate one of them~\cite{Future}.  
The main point is that proper subtraction of the box contributions 
removes all poles at finite values of $t$, so that $T_3(t)$ has poles 
only at $t=0$ and $t=\infty$, as sketched in \fig{ContourTriFigure}.
Thus we can write,
\begin{equation}
T_3(t) =   \sum_{j=-p}^p c_j  t^{j}\,.
\label{TriangleCoeffs}
\end{equation}
{}From \eqn{TripleCutMomentum} we see that the maximum power of $t$ in 
\eqn{TriangleCoeffs}, denoted by $p$, is equal to the maximum tensor 
rank encountered at the level of triangle integrals.  In a generic 
renormalizable theory such as QCD, this value is $p=3$. 

As explained in ref.~\cite{Forde}, the desired coefficient of the
triangle integral is given by $c_0$, which can be extracted by taking
the limit $t \rightarrow \infty$ and keeping only the $t^0$ contribution.
This ``Inf'' operation can be applied to either $C_3(t)$ or
the box-subtracted triple cut integrand $T_3(t)$, because the box
contributions vanish as $t\to\infty$.  In the language of OPP, the terms
with $j\neq0$ in \eqn{TriangleCoeffs} correspond to terms that integrate
to zero.   

%%% FIGURE %%%%%%%%%%%%%%%%%%%%%%%%%%%%%%%%%%%%%%%%
%%%%%%%%%%%%%%%%%%%%%%%%%%%%%%%%%%%%%%%%%%%%%%%%%%%      

We can also express the triangle coefficient using a contour integral
around $t=0$,
\begin{equation}
c_0 = {1 \over 2\pi i} \oint {d t \over t} \,  T_3(t) \,,
\end{equation}
as depicted in \fig{ContourTriFigure}.  
However, because of the special analytic form~(\ref{TriangleCoeffs})
of $T_3(t)$, it is much more efficient numerically to
evaluate this contour integral by means of a discrete Fourier projection,
\begin{equation}
c_0 = {1\over 2p+1} \sum_{j = -p}^{p} 
                        T_3\Bigl(t_0 e^{2 \pi i j/(2p + 1)}\Bigr)\,,
\end{equation}
where $t_0$ is an arbitrary complex number.
This projection removes the remaining coefficients $c_k$, $k\neq0$.
As it turns out, we do need the other coefficients in order
to subtract out triangle contributions when
evaluating bubble coefficients~\cite{Future}.  We can obtain
them from the same $2p+1$ evaluations of $T_3(t)$,
by multiplying or dividing by factors of $t$ before carrying out
the Fourier sum, 
\begin{equation}
c_k = {1\over 2p+1} \sum_{j = -p}^{p} \,
         \Bigl[ t_0 e^{2 \pi i j/(2p + 1)} \Bigr]^{-k} \, 
   T_3\Bigl(t_0 e^{2 \pi i j/(2p + 1)}\Bigr)\,.
\label{ckeqn}
\end{equation}
As we shall discuss in \sect{ResultsSection}, the
discrete Fourier projection provides excellent numerical stability.

%%%%%%%%%%%%%%%%%%%%
\subsection{Bubble Coefficients}

Next consider the bubble coefficients.  To parametrize the
remaining degrees of freedom left by the two-particle
cuts shown in \fig{TripleDoubleCutFigure}(b),
we make use of a lightlike vector $\Kf1^\mu$ constructed 
from the external momentum $K_1^\mu$ and an arbitrary lightlike
vector $\chi^\mu$.  The associated spinors are $|\Kf1^\pm\rangle$
and $|\chi^\pm\rangle$.  The normalization of 
$\chi^\mu = \sand{\chi}.{\gamma^\mu}.{\chi}/2$
is determined by the constraint that $ K_1\cdot \chi = K_1^2/2$,
which ensures that
\begin{equation}
\Kf1^\mu = K_1^\mu - \chi^\mu
\end{equation}
is lightlike.  Note that this definition of
$\Kf1$ differs from the one~(\ref{eq:def_gamma_pm}) in the triangle
discussion, and is used exclusively for the two-particle cuts associated
with the bubble coefficient.  The cut conditions $l_i^2=0$ ($i=1,2$) are
solved by the momenta,
\begin{eqnarray} 
l_i^{\mu}(y,t) &=&
 \frac{1}{2}\, K_i^\mu
 + (y-\frac{1}{2}) \left(\Kf1^\mu-\chi^\mu\right)
 + \frac{t}{2}\,\sandnm{\Kfm1}.{\gamma^\mu}.\chi
 + \frac{y(1-y)}{2\,t}\langle\chi^-|\gamma^\mu|\Kfm1 \rangle\,, \hskip .5 cm 
\label{TwoParticleParametrization}
\end{eqnarray}
with two free parameters $y$ and $t$~\cite{OPP,Forde}.  

In the two-particle cuts it is sometimes useful to restrict the cut
loop momenta to be real.  In this case, for $S_1 = K_1^2 > 0$, the cut
corresponds to a physical rescattering process.  It is convenient
to view the rescattering in the center-of-mass frame, in
which $K_1=(\sqrt{S_1},0,0,0)$, the energies of the intermediate
momenta $l_i(y,t)$ are fixed to be $\sqrt{S_1}/2$, and the phase space
can be parametrized alternatively by the polar and azimuthal angles
$\theta$ and $\phi$ for one of the two momenta, say $l_1$.  The
relation between the two parametrizations is given by,
\begin{equation}
  y = \sin^2{\theta\over2} \,, \qquad \qquad 
  t = {1\over2} \sin\theta\, e^{i\phi} \,.
\label{ytthetaphi}
\end{equation}
Then $y$ is real and restricted to $y\in [0,1]$, while
$t=\sqrt{y(1-y)}\,e^{i\phi}$ with $\phi\in [0,2\pi)$.

After subtracting box and triangle contributions from the two-particle
cut under consideration~\cite{OPP,EGK},
\begin{equation}
 C_2(y,t) \equiv A^\tree_{(1)} A^\tree_{(2)} \Bigr|_{l_i = l_i(y,t)} \,,
\label{doublecutC2}
\end{equation}
we are left with a tensorial expression $B_2(y,t)$
in terms of the loop momentum $l_i$, with maximal rank $(p-1)$.  
(In general, if the maximal rank of the triangle integrals is
$p$, the maximal rank of bubble integrals is $p-1$.)  In terms of the
parametrization (\ref{TwoParticleParametrization}), $B_2(y,t)$ is a
$(p-1)^{\rm th}$~order polynomial expression in terms of the monomials
$(1/2-y)$, $t$ and $y(1-y)/t$. The bubble coefficient is then given by the
integral~\cite{Future},
\begin{eqnarray}
b_0=\frac{1}{2\pi i}\int_0^1 dy\oint_{|t|=\sqrt{y(1-y)}}\frac{dt}{t}\,
B_2(y,t)\,.
\label{B2integral}
\end{eqnarray}
The factor of $1/t$ is a Jacobian for the change of 
variables~(\ref{ytthetaphi}) from $(\theta,\phi)$ to $(y,t)$.

As in the case of the triangle coefficients, the special analytic form
of the subtracted two-particle cut $B_2(y,t)$ allows the 
integral~(\ref{B2integral}) to be evaluated efficiently using a discrete 
Fourier projection.
Two observations are important here: the $t$ integration projects
$B_2(y,t)$ onto the terms independent of $t$, which are of maximal
power $(p-1)$ in $y$; also, the $y$ integration amounts to replacing 
positive powers of $y^n$ by rational numbers $1/(n+1)$~\cite{Forde}.
Following similar logic as in the triangle case, we can extract the 
bubble coefficient with a double discrete Fourier projection on the 
subtracted two-particle cut,
\begin{eqnarray} 
b_0=\frac{1}{(2p-1)p}\sum_{j=0}^{2(p-1)}\sum_{k=0}^{p-1}\sum_{n=0}^{p-1}\,
\frac{(y_0\, e^{2\pi ik/p})^{-n}}{n+1}\,
B_2\Bigl( y_0\, e^{2\pi i k/p} , t_0\,e^{2\pi ij/(2p-1)} \Bigr)\,,
\label{BubbleFourier}
\end{eqnarray}
where $y_0$ and $t_0$ are arbitrary complex constants.
For the case $p=3$, we use the fact that for 
$f(y) = f_0 + f_1 y + f_2 y^2$, the desired combination
$f_0+f_1/2+f_2/3$ can be written as $[f(0)+3f(2/3)]/4$.
In this way it is possible to reduce the number of values
of $y$ required, from three in \eqn{BubbleFourier} to two:
\begin{eqnarray} 
b_0=\frac{1}{20}\sum_{j=0}^{4}
\biggl[  B_2\Bigl( 0, t_0\,e^{2\pi ij/5} \Bigr)
+ 3 B_2\Bigl( 2/3, t_0\,e^{2\pi ij/5} \Bigr) \biggr]\,.
\label{AltBubbleFourier}
\end{eqnarray}
One can also reduce the number of values of $t$ sampled,
from five down to three or four, using lower-order roots of unity 
(independently of how $y$ is treated).
In a similar fashion to \eqn{ckeqn}, higher-rank tensor bubble coefficients 
may be extracted by weighting the sum~(\ref{BubbleFourier}) differently.
(Such coefficients would feed into the calculation of tadpole
coefficients.  They are not needed for the case of massless internal lines
treated in this paper.)

Due to the physical interpretation of the two-particle cut as a
rescattering, with real intermediate momenta living on a sphere,
an alternative projection formula from
\eqns{BubbleFourier}{AltBubbleFourier} may be found in terms of 
spherical harmonics $Y_{l,m}(\theta,\phi)$.  To do so we
change from the variables $y$ and $t$ to the spherical coordinates 
$\theta$ and $\phi$ via \eqn{ytthetaphi}.  In these variables,
the loop momentum~(\ref{TwoParticleParametrization}) is linear in the 
spherical harmonics $Y_{l,m}$ with $l=1$ and $m=0,\pm1$, because
\begin{eqnarray} 
{1\over2}-y &=& {1\over2} \cos\theta 
             = \sqrt{\pi \over 3} \,  Y_{1,0}(\theta,\phi)\,, \nn \\
t &=& {1\over2} \sin\theta \,  e^{i\phi} 
   = - \sqrt{2 \pi \over 3}\, Y_{1,1}(\theta,\phi)\,, \nn \\
\frac{y(1-y)}{t} &=& {1\over2} \sin\theta \, e^{-i\phi} 
   =  \sqrt{2 \pi \over 3}\, Y_{1,-1}(\theta,\phi)\,.
\end{eqnarray}
The two-particle cut with box and triangle contributions subtracted is
then a superposition of spherical harmonics,
\begin{eqnarray} 
B_2(\theta,\phi) = \sum_{|m|\le l\le p-1} b_{l,m} \, Y_{l,m}(\theta,\phi)\,.
\end{eqnarray}
The scalar bubble coefficient is just $b_{0,0}$, up to a normalization 
constant.  Using \eqn{TwoParticleParametrization}, the higher
spherical-harmonic coefficients $b_{l,m}$ can be related to the 
coefficients of the higher-rank tensor integrals.

%%%%%%%%%%%%%%%%%%%%%%%%%%%%%%%%%%%%%%%%%%%%%%%%%%%%%
\section{Rational Contributions}
\label{RationalSection}

%%% FIGURE %%%%%%%%%%%%%%%%%%%%%%%%%%%%%%%%%%%%%%%%
\begin{figure}
\begin{center}
\includegraphics[width=.3\textwidth]{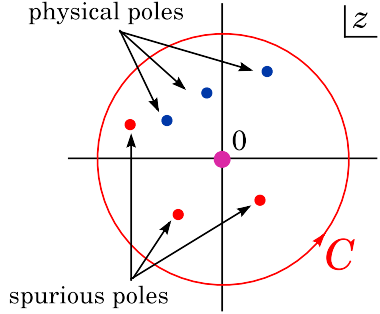}
%\centerline{\epsfxsize 2 truein \epsfbox{LoopCauchy.eps}}
 \caption{Using Cauchy's theorem, rational terms in loop
amplitudes can be reconstructed from residues at poles in the 
complex plane.  The poles are of two types: physical and spurious.
All pole locations are known {\it a priori}.
Residues at physical poles are obtained from on-shell recursion.
Residues at spurious poles are obtained from the cut parts.}
\label{LoopCauchyFigure}
\end{center}
\end{figure}
%%%%%%%%%%%%%%%%%%%%%%%%%%%%%%%%%%%%%%%%%%%%%%%%%%%

We now turn to the question of computing the rational terms $R_n$ in 
the amplitude (\ref{CutRational}).  Here we use the on-shell 
recursive approach for one-loop amplitudes~\cite{Bootstrap,Genhel}, 
modifying it to make it more amenable to numerical evaluation in an
automated program.  As is true for the cut parts, an important 
feature of on-shell recursion is that it displays a modest growth
in computational resource requirements --- compared to the rapid growth 
with a traditional Feynman-diagram approach --- as the number of 
external particles increases.  

At one loop, as at tree level, on-shell recursion provides a
systematic means of determining rational functions, using knowledge of
their poles and residues.  At loop level, however, a number
of new issues must be addressed, including the appearance of
branch cuts, spurious singularities, and the behavior of loop
amplitudes under large complex deformations.  In some cases, ``unreal
poles'' develop~\cite{DoublePole}, which are poles present with
complex but not real momenta.  The appearance of branch cuts does not
present any difficulties because we use on-shell recursion only for
the cut-free rational remainders $R_n$.  As noted in ref.~\cite{Genhel}, 
we can sidestep the problems of unreal poles by choosing appropriate
shifts within the class given below in \eqn{MomShift}.  Finally, 
we may determine the behavior of amplitudes under large complex 
deformations by using auxiliary recursion relations.

\subsection{General Principles}

On-shell recursion relations may be derived by considering
deformations of amplitudes characterized by a single complex parameter
$z$, such that all external momenta are left on shell~\cite{BCFW}.  In
the massless case, it is particularly convenient to shift the momenta
of two external legs, say $j$ and $l$,
\begin{eqnarray}
&k_j^\mu &\rightarrow k_j^\mu(z) = k_j^\mu - 
      {z\over2}{\langle{j^-}|{\gamma^\mu}|{l^-}\rangle} \,, \nonumber\\
&k_l^\mu &\rightarrow k_l^\mu(z) = k_l^\mu + 
      {z\over2}{\langle{j^-}|{\gamma^\mu}|{l^-} \rangle} \,.
\label{MomShift}
\end{eqnarray}
We denote the shift in eq.~(\ref{MomShift}) as
a $[j, l \rangle$ shift.  This shift has the required property that the
momentum conservation is left undisturbed, while shifted momenta are
left on-shell, $ k_j^2(z) = k_l^2(z) = 0$.

On-shell recursion relations follow from evaluating the contour integral,
\begin{equation}
{1\over 2\pi i} \oint_C dz \,{R_n(z)\over z} \,,
\end{equation}
where the contour is taken around the circle at infinity, as depicted
in fig.~\ref{LoopCauchyFigure}, and $R_n(z)$ is $R_n$ evaluated at the
shifted momenta~(\ref{MomShift}).  If the rational terms under
consideration vanish as $z\rightarrow \infty$, the contour integral
vanishes and Cauchy's theorem gives us a relationship between the
desired rational contributions at $z=0$, and a sum over residues of
the poles of $R_n(z)$, located at $z_{\alpha}$,
\begin{equation}
R_n(0) = -\sum_{{\rm poles}\ \alpha} \Res_{z=z_\alpha}  
{R_n(z)\over z} \,.
\label{ResidueSum}
\end{equation}
On the other hand, if the amplitude does not vanish as $z\rightarrow
\infty$, there are additional contributions, which we can
obtain from an auxiliary recursion relation~\cite{Genhel}.

%%% FIGURE %%%%%%%%%%%%%%%%%%%%%%%%%%%%%%%%%%%%%%%%%%%%%%%%%%%%%%%%%
\begin{figure}
\begin{center}
\includegraphics[width=.8\textwidth]{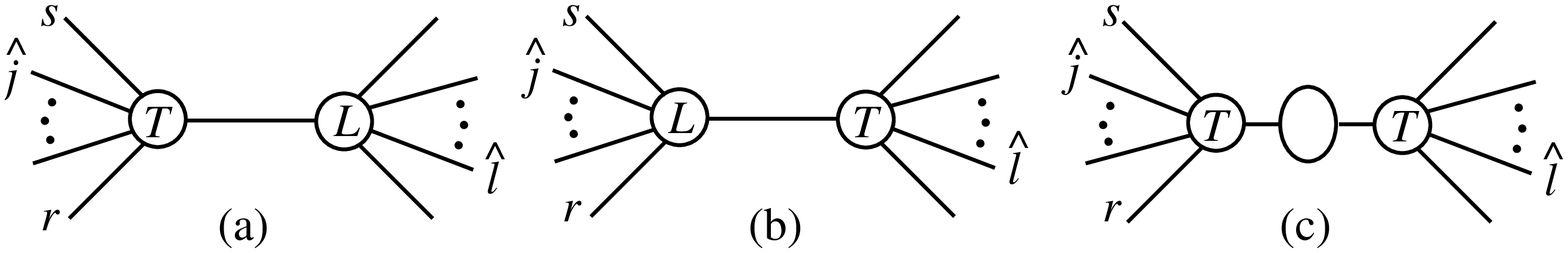}
 \caption{Diagrammatic contributions to on-shell recursion at one-loop
for a $[j, l \rangle$ shift.  The labels ``$T$'' and ``$L$'' refer
respectively to (lower-point) tree amplitudes $A^\tree$
and rational parts of one-loop amplitudes $R$.  
The central blob in (c) is the rational part of a one-loop 
factorization function ${\cal F}$~\cite{BernChalmers}.}
\label{LoopGenericFigure}
\end{center}
\end{figure}
%%%%%%%%%%%%%%%%%%%%%%%%%%%%%%%%%%%%%%%%%%%%%%%%%%%%%%%%%%%%%%%%%%%%

Poles in the $z$-shifted one-loop rational terms,
labeled by $\alpha$ in \eqn{ResidueSum},
may be separated into two classes
as shown in fig.~\ref{LoopCauchyFigure}: physical and spurious.
The physical poles are present in the full amplitude $A_n$, and
correspond to genuine, physical factorization poles 
(collinear or multiparticle).  
The spurious poles are not poles of $A_n$; they cancel between
the cut parts $C_n$ and rational parts $R_n$.  They
arise from the presence of tensor integrals in the underlying
field-theory representation of the amplitude.  
Our method avoids the need to perform the reduction of such tensor
integrals explicitly, because of the use of a basis of master integrals.
The reduction happens implicitly, and leaves its trace in the presence
of Gram determinant denominators.  These denominators
give rise to spurious singularities in individual terms.
Separating the different contributions, we may write,
\begin{equation}
R_n(z) = R^D_n(z) + R^S_n(z) + R^{{\rm large}\ z}_n(z)\,, 
\label{totrec}
\end{equation}
where $R^D_n$ contains all contributions from physical poles, $R^S_n$ the
contributions from spurious poles, and $R^{{\rm large}\ z}_n$ the possible
contributions from large deformation parameter $z$, 
if $R_n(z)$ does not vanish there.
More explicitly, from elementary complex variable theory, under
the shift (\ref{MomShift}) the rational terms can be expressed as a
sum over pole terms and possibly a polynomial in $z$,
\begin{eqnarray}
R_n^D(z) &=& \sum_\alpha {A_\alpha \over z - z_\alpha}\,,
\hskip 2 cm 
R^S_n(z)  =  \sum_\beta \Biggl({B_\beta \over (z - z_\beta)^2} +
                             {C_\beta \over z - z_\beta } \Biggr)\,, \nn\\ 
&& \hskip 2 cm 
R^{{\rm large}\ z}_n(z) = \sum_{\sigma=0}^{\sigma_{\rm max}}
                                         D_\sigma z^\sigma \,,
\label{RnShiftGeneric}
\end{eqnarray}
where the coefficients $A_\alpha, B_\beta, C_\beta, D_\sigma$ are
functions of the external momenta. The poles in $z$ in
\eqn{RnShiftGeneric} are shown in \fig{LoopCauchyFigure}.
The physical poles labeled by $\alpha$ are
generically single poles. (Some shift choices may lead to double
poles~\cite{DoublePole}; we can generally avoid such 
shifts~\cite{Genhel}.)  In general, in a renormalizable
gauge theory, the spurious poles, labeled by $\beta$, may be either
single or double poles~\cite{Future}.  If $R_n(z)$ vanishes for large
$z$, the $D_\sigma$ are all zero.  If not, then $D_0$ gives a
contribution to the physical rational terms, $R_n(0)$.

The contributions of the physical poles may be obtained efficiently
using the on-shell recursive terms represented by the diagrams in
\fig{LoopGenericFigure}.  The tree ``vertices'' labeled by ``$T$''
denote tree-level on-shell amplitudes $A_m^\tree$, while the loop vertices
``$L$''are the rational parts of on-shell (lower-point) one-loop
amplitudes $R_m$, $m<n$, as defined in \eqn{RatDef}.
The contribution in \fig{LoopGenericFigure}(c) involves the rational
part of the additional factorization function ${\cal F}$~\cite{BernChalmers}.
It only appears in multi-particle channels, and only if the tree amplitude 
contains a pole in that channel. 
Each diagram is associated with a physical pole in the $z$ plane, 
illustrated in \fig{LoopCauchyFigure}, whose location is given by,
\begin{equation}
z_{\alpha} = z_{rs} \equiv
{K_{r\cdots s}^2 \over \langle{j^-}| \; {\slash \hskip -.3 cm 
     K_{r\cdots s}} \, |{l^-}\rangle } \,,
\end{equation}
where $K_{r\ldots s} = k_{r} + k_{r+1} + \cdots + k_s$.
This pole arises from the vanishing of shifted propagators, 
$K_{r\ldots s}^2(z_{rs}) = 0$.  Generically the sum
over $\alpha$ is replaced by a
double sum over $r,s$, labeling the recursive diagrams, where legs
labeled $\hat\jmath$ and $\hat l$ always appear on opposite sides
of the propagator in \fig{LoopGenericFigure}.
The computation of the recursive diagrams has been described 
in refs.~\cite{Bootstrap,OneLoopMHV,OnShellReview}, to which we refer 
the reader for further details.

What about the contributions of the spurious poles?
One approach is to find a ``cut completion''~\cite{Bootstrap,Genhel},
which is designed by adding appropriate rational terms to $C_n$
in order to cancel entirely the spurious poles in $z$ within the 
redefined cut terms $\hat{C}_n$.
Because the complete amplitude is free of the spurious poles,
this procedure ensures that the redefined rational terms $\hat{R}_n$
are free of them.  The cut completion makes it unnecessary to
compute residues of spurious poles (although additional ``overlap''
diagrams are introduced).  It is very helpful for deriving compact
analytic expressions for the amplitudes.  This approach 
has led to the computation of the rational terms
for a variety of one-loop MHV amplitudes with an arbitrary number of
external legs~\cite{Bootstrap,OtherBootstrap,Genhel}, as well
as for six-point amplitudes.  In general, it
should be possible to construct a set of cut completions using integral
functions of the type given in ref.~\cite{CGM} to absorb spurious
singularities.

For the purposes of a numerical program, however, it is simpler to
extract the spurious residues from the known cut parts.  These residues
are guaranteed to be the negatives of the spurious-pole residues in 
the rational part.  That is, the spurious contributions are,
\begin{equation}
R^S_n(0) =
 -\sum_{{\rm spur.\ poles}\ \beta} \Res_{z=z_\beta} {R_n(z)\over z}
= \sum_{{\rm spur.\ poles}\ \beta} \Res_{z=z_\beta} {C_n(z)\over z} \,,
\label{spureqn}
\end{equation}
where $C_n(z)$ is the shifted cut part appearing in \eqn{CutRational}.
The spurious poles $\beta$ correspond to the vanishing of shifted Gram
determinants, $\Delta_m(z)=0$ for $m=2,3,4$, associated with 
bubble, triangle and box integrals.  
(In the case of massless internal propagators, the bubble Gram 
determinant does not generate any spurious poles.)  

A simple example of a spurious singularity in the cut 
part~(\ref{IntegralBasis}) is from a bubble term of the form,
\begin{equation}
b_i \, I_2^i 
=  { \hat{b}_i  \over (K_1^2-K_2^2)^2 } \ln(-K_1^2) + \cdots,
\label{spurexample}
\end{equation}
where $\hat{b}_i$ is smooth as $K_1^2 \to K_2^2$, and $K_1+K_2+k_3=0$
for some massless momentum $k_3$.  The denominator factor
$(K_1^2-K_2^2)$ is the square root of the Gram determinant for 
a triangle integral with two massive legs,
$K_1$ and $K_2$, and one massless leg, $k_3$.  
Under the $[j, l \rangle$ shift, there will be a 
value of $z$, $z_\beta$, for which the shifted denominator vanishes linearly,
$K_1^2(z) - K_2^2(z) \sim z-z_\beta$ 
(unless $j$ and $l$ both belong to the same massive momentum cluster, 
$K_1$ or $K_2$, in which case the Gram determinant is unshifted).
{}From \eqn{spureqn} we see that we
only need the rational pieces of the spurious-pole residues of the cut
part, because $R^S_n(0)$ is rational.
From \eqn{spurexample}, we see that there can only be a rational
piece if we have to series expand the logarithm to compute the residue.
Hence the spurious pole in the bubble coefficient $b_i$ 
must be of at least second order in $(K_1^2-K_2^2)$.
At order $\e^0$, box and triangle integrals contain dilogarithms
and squared logarithms, which must be expanded to second order
to obtain a rational piece.  Thus the spurious poles of box and triangle
coefficients must be at least of third order for rational terms 
to be generated.

To extract a residue from $C_n(z)/z$, we evaluate the integral coefficients 
$d_i,c_i,b_i$ numerically for complex, shifted momenta 
in the vicinity of the spurious pole, using our implementation of 
the results of~\sect{CutsSection}.  We also need to evaluate the loop
integrals.  First, however, we perform an analytic series 
expansion of the integrals around the vanishing Gram determinants.  
For example, the three-mass triangle integral, 
$I_3^{\rm 3m}(s_1,s_2,s_3)$, close to the surface of its vanishing
Gram determinant,
\begin{equation}
\Delta_3 \equiv s_1^2+s_2^2+s_3^2-2s_1 s_2-2s_1 s_3-2s_2 s_3\ \to\ 0 \,,
\label{Delta3mdef}
\end{equation}
behaves as,
\begin{eqnarray}
I_3^{\rm3m}(s_1,s_2,s_3) &\rightarrow&
 - {1\over2} \sum_{i=1}^3 \ln(-s_i)
     {s_i-s_{i+1}-s_{i-1}\over s_{i+1} s_{i-1}}
     \left[ 1 - {1\over6} {\Delta_3\over s_{i+1} s_{i-1}}  
 + {1\over30} \left( {\Delta_3\over s_{i+1} s_{i-1}} \right)^2\right] \nn \\
&&\hskip0.1cm \null
+ {1\over6} {\Delta_3\over s_1s_2s_3} - {s_1+s_2+s_3\over120}
      \left({\Delta_3\over s_1s_2s_3} \right)^2+ \cdots \,,
\label{tri3mexp}
\end{eqnarray}
where the index $i$ on the shifted invariant, $s_i \equiv s_i(z)$, 
is defined mod 3.
In this expression the logarithms are to be expanded according to, 
\begin{equation}
\ln(-s) \rightarrow {s-s_\beta \over s_\beta} - {1\over 2}
   {(s-s_\beta)^2 \over s_\beta^2} + \cdots \,,
\label{LogExpand}
\end{equation}
where $s=s(z)$, and $s_\beta=s(z_\beta)$ is the value of the shifted
invariant at the location $z_\beta$ of the spurious pole.  
The leading order of \eqn{tri3mexp} matches the expansion found 
in ref.~\cite{CGM}.  
In the integral expansions we need keep only rational terms,
including terms that can become rational after further series
expansion around a generic point, such as \eqn{LogExpand}.  
Thus we may avoid computing any logarithms or polylogarithms at 
complex momentum values.  The expression obtained by replacing $C_n(z)$ 
according to these rules, in the vicinity of $z_\beta$, will be
denoted by $E_n^\beta(z)$.
In ref.~\cite{Future} we present the complete set of integral expansions
needed in the calculations, as well as a convenient method for
generating them from a dimension-shifting formula~\cite{BDKIntegrals}.

\subsection{Discrete Fourier Sum for Spurious Residues}

%%% FIGURE %%%%%%%%%%%%%%%%%%%%%%%%%%%%%%%%%%%%%%%%
\begin{figure}
\begin{center}
\includegraphics*[width=0.4\textwidth]{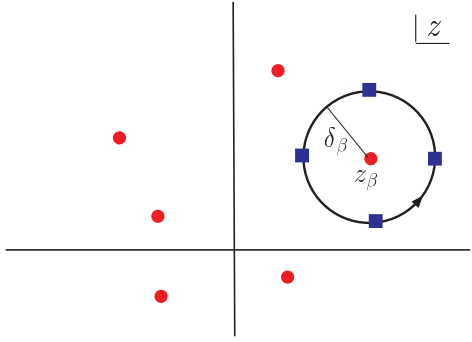}
%\centerline{\epsfxsize 2.5 truein \epsfbox{Contour.eps}}
 \caption{We obtain the residue at the spurious pole located at 
$z = z_\beta$ in the complex $z$ plane by a discrete Fourier sum,
evaluating $E_n^\beta(z)$ on the (blue) squares on the circle of radius
$\delta_\beta$ centered on $z_\beta$.  In this figure four points
are shown, although in practice we use ten points. 
The locations of other poles are represented by (red) dots. 
We ensure that $\delta_\beta$ is sufficiently small so that other 
poles give a negligible contribution to the residue.}
\label{ContourRatFigure}
\end{center}
\end{figure}

%%%%%%%%%%%%%%%%%%%%%%%%%%%%%%%%%%%%%%%%%%%%%%%%%%%      

Similarly to the case of triangle and bubble coefficients, 
we extract each required spurious-pole residue from the cut parts
by using a discrete Fourier sum.  We evaluate $E_n^\beta(z)$ at 
$m$ points equally spaced around a circle of radius $\delta_\beta$ 
in the $z$ plane, centered on the pole location $z_\beta$, 
as depicted in \fig{ContourRatFigure}; {\it i.e.},
$z = z_\beta + \delta_\beta e^{2\pi ij/m}$, for $j=1,2,\ldots,m$.
In contrast to the $t$-plane analysis used earlier to obtain 
triangle coefficients, however, we do not know the residues 
at other poles {\it a priori\/}, so we cannot subtract them easily.  
(Indeed, the function $E_n^\beta(z)$ 
we are analyzing is only rational in the vicinity of $z_\beta$, due
to our use of the rational parts of the integral expansions around
this point.)
Here the discrete Fourier sum is an approximation to the contour
integral, whereas in the previous section it was exact.
We can make the approximation arbitrarily accurate in principle,
by choosing $\delta_\beta$ to be arbitrarily small.  With finite 
precision, however, numerical round-off error forces us to work at 
finite $\delta_\beta$.  When extracting the residue
of a spurious pole we must also ensure that there are no other poles
inside or near the circle.   To obtain the contributions of the 
spurious poles to $R_n(0)$ in \eqn{spureqn} we evaluate,
\begin{equation}
R^S_n(0)\ \simeq\ 
 {1\over m} \sum_\beta \sum_{j = 1}^{m}  \delta_\beta e^{2 \pi ij /m}
 {E_n^\beta(z_\beta  + \delta_\beta e^{2 \pi i j/m}) \over z_\beta  + 
         \delta_\beta e^{2 \pi i j/m} } \,.
\end{equation}
The sum over $\beta$ runs over the location of
all spurious Gram determinant poles that contribute to rational terms.
Equivalently, we can extract the coefficients $B_\beta$ and 
$C_\beta$ in \eqn{RnShiftGeneric} via,
\begin{eqnarray}
B_\beta &\simeq& -{1\over m} \sum_{j = 1}^{m} 
          \Bigl[ \delta_\beta  \, e^{2 \pi ij /m} \Bigr]^2
          E_n^\beta(z_\beta+\delta_\beta e^{2 \pi i j/m})\,, \nn \\
\hskip 1 cm 
C_\beta &\simeq& -{1\over m} 
        \sum_{j = 1}^{m} \delta_\beta \, e^{2 \pi ij /m} \,
          E_n^\beta(z_\beta+\delta_\beta e^{2 \pi i j/m})\,.
\end{eqnarray}
For the results presented in the next section we choose $m=10$ points
in the discrete sum.  In general, an increase in $m$ increases
the precision, but at the cost of computation time.

We choose $\delta_\beta$ to be much smaller than the distance to
nearby poles, but not so small as to lose numerical precision.
Typically at ``standard'' double precision we use a value of
$\delta_\beta = 10^{-2}$.  If the contributions from the nearby poles
are unusually large, then we find a large variation in the absolute
value of each term in the sum.  If this happens we reduce
$\delta_\beta$ until either the variation is acceptable, or we cross a
minimum value of $\delta_\beta$, beyond which the point becomes
unstable because of round-off error. We deal with such points as
described below.

\subsection{Numerical Stability}

In addition to the value of $\delta_\beta$ becoming too small,
other cancellations can also sometimes cause a loss of precision,
giving rise to a potentially unstable kinematic point.
In order to identify such phase-space points more generally,
we apply consistency checks independently to the cut and rational
parts of the amplitude.  For the cut part we test how well the 
known, non-logarithmic $1/\e$ singularities are reproduced.  
Because the only source of such $1/\e$ poles are the bubble integrals, 
for the $n$-gluon amplitudes, for example, we have~\cite{GG,KST},
\begin{equation}
A_n^{\oneloop}|_{1/\e,\,{\rm non-log}}
= {1\over \e} \sum_k b_k 
= - \left[\frac{1}{\e}
\left(\frac{11}{3}-\frac{2}{3}\frac{n_{\! f}}{N_c}\right)\right]
A_n^{\rm tree}
\,,
\label{BubbleSingularity}
 \end{equation}
where $n_{\!f}$ is the number of quark flavors and the sum on $k$ runs
over all bubble integrals.  As a practical matter it is sufficient to
check that the divergent term divided by the tree amplitude is real.
(For helicity configurations with vanishing tree amplitudes the
cut contributions vanish, so no check is required.)  Because
bubble coefficients are computed from expressions where
triangle and box contributions have been subtracted, any instabilities
in the latter are also detected with this $1/\eps$ consistency check.

In general this test is not sufficient for finding all the unstable
points of the full amplitude, because some of the instability comes from
computing the spurious residues for rational terms.  A related test,
which suffices to find all remaining instabilities, comes from the
requirement that each spurious singularity must cancel in the sum
over bubble coefficients.  This cancellation can be understood by 
applying the $[j, l \rangle$ shift to \eqn{BubbleSingularity},
and making use of the fact that $A_n^\tree$ has no spurious poles.
For each spurious-pole residue that contributes to the rational part,
we therefore check that the sum of discrete Fourier sums 
over all bubble coefficients,
\begin{equation}
\sum_k \sum_{j=1}^m \delta_\beta e^{2\pi ij/m} \,
 { b_k(z_\beta + \delta_\beta e^{2\pi ij/m}) } \,, 
\label{littleIR}
\end{equation}
vanishes to within a specified tolerance.

If a phase-space point fails the above stability conditions we
recalculate the point in a manner that improves its stability. Various
strategies have been proposed in the literature to handle unstable
points.  One approach is to modify the standard integral basis
(\ref{IntegralBasis}) so as to absorb the Gram determinant
singularities into well-defined functions~\cite{FiveGluon, Zqqgg, CGM,
Denner, Binoth}. This approach is related to using a cut
completion~\cite{Bootstrap}.  Other approaches are to interpolate
across the singular region or to series expand the integrals in the
singular region~\cite{GramInterpolate,Denner}.  A third approach is to
simply redo unstable points at higher precision, {\it e.g.} as in
ref.~\cite{CutTools}.

We have found the high-precision approach to be effective for
eliminating the remaining instabilities in our program.  It is robust
and simple to implement; a detailed analysis of the instabilities is
not needed, and we can use the standard basis of integrals with no
interpolations or expansions of the integrals around unstable points.
Our implementation of on-shell methods already has only a small
fraction of unstable phase-space points; hence the overhead of
recomputing them at higher precision is relatively small.  We use the
{\tt QD} package~\cite{QD}, switching to ``double-double'' precision,
that is approximately 32 decimal digits. If the stability test were to
fail at this level of precision, we switch to ``quadruple-double''
precision, corresponding to approximately 64 digits of precision; for
all amplitudes calculated here, this happens rarely, if ever.  To
compute the integrals to higher precision, we implement the
polylogarithms which enter the integrals using a series expansion
to a sufficiently high order.  If the $1/\ep$ test
(\ref{BubbleSingularity}) fails then we recompute the entire cut part
at higher precision, but if the spurious-pole test~(\ref{littleIR}) fails we
only recompute those pieces containing unstable Gram determinant
singularities.

Further details, as well as all integral expansions used to extract the 
spurious residues from the cut part, will be given elsewhere~\cite{Future}.

%%%%%%%%%%%%%%%%%%%%%%%%%%%%%%%%%%%%%%%%%%%%%%%%%%%%%
\section{Results}
\label{ResultsSection}

We now discuss the numerical stability of our implementation. 
Our stability tests use sets of 100,000 points for 
$2 \rightarrow (n-2)$ gluon scattering, generated with a flat 
phase-space distribution using the {\tt RAMBO}~\cite{RAMBO} algorithm.  
We impose kinematic cuts on the outgoing gluons, following ref.~\cite{EGK}:
\begin{equation}
E_T > 0.01 \sqrt{s}\,, \hskip 1 cm \eta<3\,, \hskip 1 cm \Delta_R >0.4\,,
\label{KinematicCuts}
\end{equation}
where $E_T$ is the gluon transverse energy, $\eta$ is the pseudorapidity, 
and $\Delta_R = \sqrt{\Delta_\eta^2 + \Delta_\phi^2}$ is the separation 
cut between pairs of gluons.
The center-of-mass energy $\sqrt{s}$ is chosen to be $2$ TeV and the
scale parameter $\mu$ (arising from divergent loop integrals) is set 
to $1$ TeV.

We computed one-loop six-, seven- and eight-gluon amplitudes 
for $n_{\!f}=0$ with \BlackHat{} at each phase-space point,
and compared the output against a target expression, obtained either from
known analytic results, or from {\tt BlackHat} itself
using quadruple-double precision ($\sim \! 64$ digits).
As an additional test, we also used ordinary double
precision to compare to the numerical results of refs.~\cite{EGZ,GKM}
at the quoted phase-space points.  We find agreement for the five- and
six-gluon amplitudes for all helicity configurations, to within their
quoted accuracy, after accounting for external phase conventions and
the incoming-particle convention implicitly used in ref.~\cite{GKM}.
We also find agreement with the numerical results of
ref.~\cite{Genhel} at the quoted phase-space points for the six-,
seven- and eight-point maximally helicity violating (MHV) amplitudes
presented here.

%%% FIGURE %%%%%%%%%%%%%%%%%%%%%%%%%%%%%%%%%%%%%%%%
\begin{figure}
\begin{center}
\includegraphics[width=.97 \textwidth,clip]{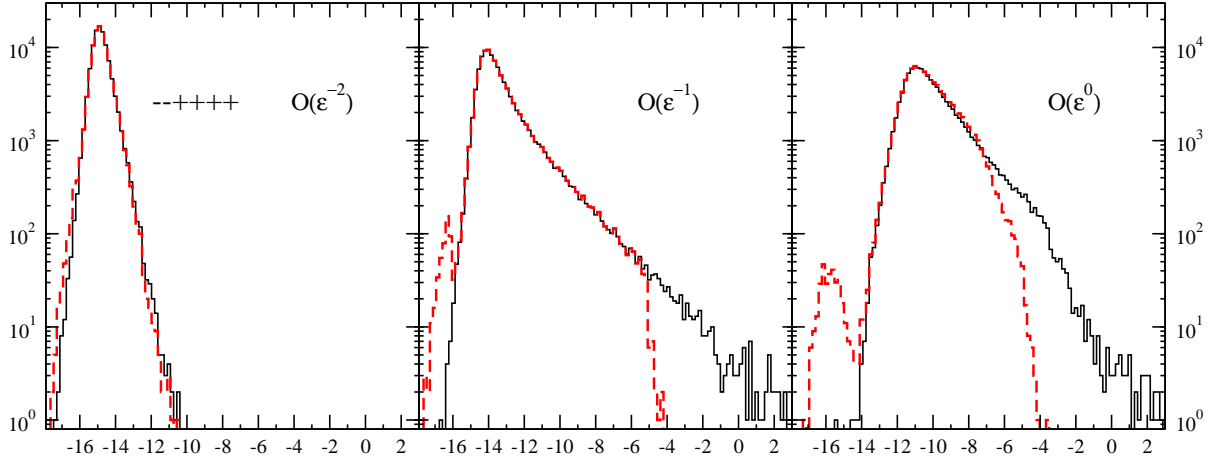}
\caption{The distribution of the logarithm of the relative error for
100,000 phase-space points in the $1/\ep^2$, $1/\ep$ and finite ($\e^0$) 
components of the 
six-point MHV amplitude $A_6(1^-,2^-,3^+,4^+,5^+,6^+)$. The solid
(black) curve shows the distribution run entirely with ordinary double
precision, and the dashed (red) curve shows it when contributions
identified as unstable --- following the discussion
of~\sect{RationalSection} --- are evaluated using higher
precision. The target values use analytic results from
refs.~\cite{UnitarityMethod, Fusing, Bootstrap}. }
\label{AmmppppTailFigure}
\end{center}
\end{figure}
%%%%%%%%%%%%%%%%%%%%%%%%%%%%%%%%%%%%%%%%%%%%%%%%%%%      

\begin{figure}
\begin{center}
\includegraphics[width=.97\textwidth,clip]{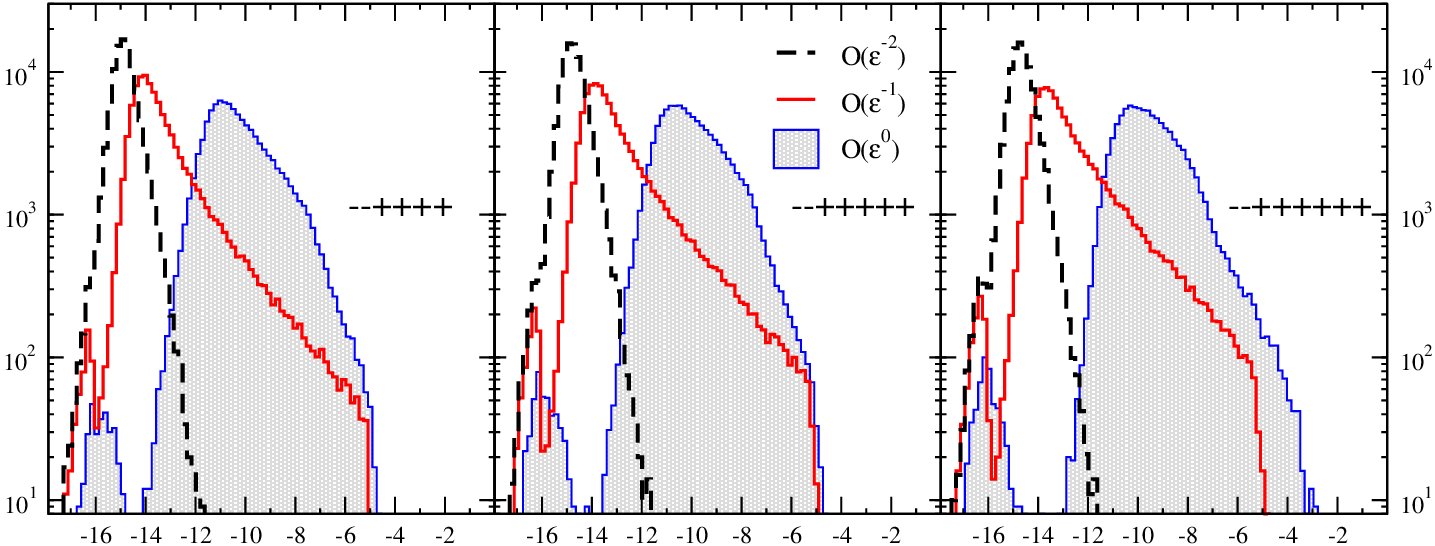}
\caption{The distribution of the logarithm of the relative error 
over 100,000 phase-space points for the MHV amplitudes
$A_6(1^-,2^-,3^+,4^+,5^+,6^+)$, $A_7(1^-,2^-,3^+,4^+,5^+,6^+,7^+)$ and
$A_8(1^-,2^-,3^+,4^+,5^+,6^+,7^+,8^+)$. The dashed (black) curve 
in each histogram gives the relative error 
for the $1/\e^2$ part,  the solid (red) curve gives the $1/\e$ 
singularity, and the shaded (blue) distribution gives 
the finite $\ep^0$ component of the corresponding helicity amplitude.  
The target expression is computed from an analytic
formula~\cite{UnitarityMethod, Fusing, Bootstrap, OneLoopMHV}.}
\label{fig:splitMHVs}
\end{center}
\end{figure}

\begin{figure}
\begin{center}
\includegraphics[width=.97\textwidth,clip]{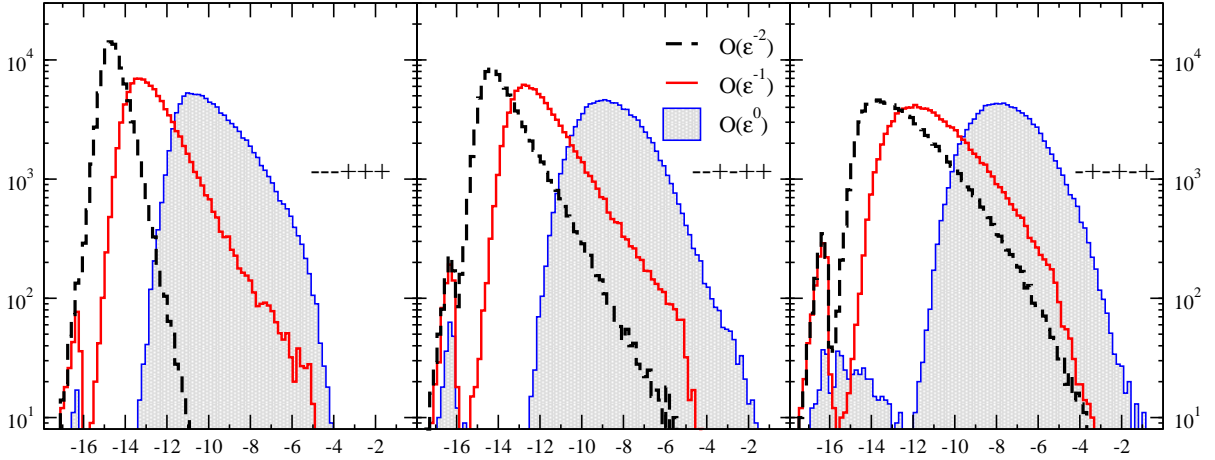}
%\centerline{\epsfxsize 2.5 truein \epsfbox{NMHVs_6pt.eps}}
\caption{The distribution of the logarithm of the relative error
for the six-point NMHV amplitudes
$A_6(1^-,2^-,3^-,4^+,5^+,6^+)$, $A_6(1^-,2^-,3^+,4^-,5^+,6^+)$ and
$A_6(1^-,2^+,3^-,4^+,5^-,6^+)$.
The dashed (black) curve in each histogram gives the relative error 
for the $1/\e^2$ part, the solid (red) curve gives the $1/\e$ 
singularity, and the shaded (blue) curve gives 
the finite $\ep^0$ component of the corresponding amplitude.
The target expression is a quadruple-double-precision \BlackHat{}
evaluation.}
\label{fig:NMHVs}
\end{center}
\end{figure}

The histograms in figs.~\ref{AmmppppTailFigure}--\ref{fig:NMHVs} show the
results of our study of numerical precision.  For these plots, the
horizontal axis is the logarithmic relative error,
\begin{equation}
\log_{10}\left(\frac{|A_n^{\rm num}-A_n^{\rm target}|}
                                {|A_n^{\rm target}|}\right)\,,
\end{equation}
for each of the $1/\e^2$, $1/\e$ and $\e^0$ components of the
numerical amplitude $A_n^{\rm num}$ obtained from \BlackHat{}.  The
vertical axis in these plots shows the number of phase-space
points in a bin that agree with the target to a specified
relative precision.  We use a logarithmic vertical scale to visually
enhance the tail of the distribution, so as to illustrate the
numerical stability.

For the MHV amplitudes in figs.~\ref{AmmppppTailFigure}
and~\ref{fig:splitMHVs}, we used analytic expressions from
refs.~\cite{UnitarityMethod,Fusing,Bootstrap,OneLoopMHV} as the target
expressions $A_n^{\rm target}$.  For the next-to-MHV (NMHV) amplitudes, 
analytic expressions are available~\cite{Fusing,
BCFCutConstructible,XYZ,Genhel}, although for \fig{fig:NMHVs}, we 
generated the target with \BlackHat{}, using
quadruple-double precision. 
This is more than sufficient to ensure numerical stability in target
expressions for the purposes of the comparison.
We note that the ability to switch easily to higher precision 
is quite helpful in assessing numerical stability in
any new calculation.

First consider the MHV six-point amplitude $A_6(1^-,2^-,3^+,4^+,5^+,6^+)$.
\Fig{AmmppppTailFigure} illustrates the numerical
stability of \BlackHat{} for this amplitude, with and without the use
of higher precision on the points identified as unstable.  The plots show
the distribution of relative errors for the $1/\ep^2$, $1/\ep$ and
$\ep^0$ components over 100,000 phase-space points.  The $1/\ep^2$
distribution has extremely small errors, peaking at a relative error
of nearly $10^{-15}$, while the right-side
tail falls rapidly.  For the $1/\e$ and finite $\e^0$ components the
peaks shift to the right, to a relative precision of around $10^{-14}$
and $10^{-11}$, and fall less steeply.  This feature is not surprising, 
because of the larger number of computational steps needed
for these parts of the amplitudes: for $1/\e^2$ terms, only box
coefficients contribute (for this helicity pattern triangle
integrals do not appear); for the $1/\e$ contribution, bubble
coefficients contribute too; for the finite part, rational
terms contribute as well.  As one proceeds from box to triangle, bubble, 
and then to rational terms, each step relies on previous steps, and so
numerical errors accumulate.  

In each plot in \fig{AmmppppTailFigure}
the solid (black) curve corresponds to the exclusive use of ordinary
double precision (16 decimal digits), showing good
stability for the raw algorithm for all three components.  The dashed
(red) curve shows the effect of turning on higher precision for
contributions identified as unstable, using the criteria discussed
in~\sect{RationalSection}.  This completely suppresses the
already-small tail above a relative error of about $10^{-5}$. The
points populating the right-hand tail in the ordinary double precision
calculation, displayed in the solid (black) curve, then move to the left in
the dashed (red) curve, giving rise to a secondary peak around
a relative error of machine precision, or $10^{-16}$.  (The
comparison with the target is performed in ordinary double-precision, 
even though higher precision is used in intermediate steps.)  
This twin-peak feature is visible in the $1/\eps$ and $\eps^0$ components.  
It is due our recalculation of the
entire cut part, at higher precision, whenever a phase-space
point fails the $1/\eps$ consistency check~(\ref{BubbleSingularity}).
When the spurious-pole stability test~(\ref{littleIR}) fails, 
the point generally falls to the right of the secondary peak,
because we only recalculate those pieces that contain
the unstable spurious singularity.

Another important feature that can be observed in
\fig{AmmppppTailFigure} is that the ``effective cutoff'' is sharp: for
the $\eps^0$ terms almost no points below $10^{-5}$ are identified as
unstable.
In a practical calculation,
given Monte-Carlo integration errors and other uncertainties, a cutoff
in the relative error of $10^{-5}$ is overly stringent. It does,
however, illustrate the control over instabilities achieved
in \BlackHat, which becomes more important for more complicated
processes.  It is interesting to note that modest additional
computation time is required to achieve a cutoff of $10^{-5}$,
compared to, say, $10^{-2}$.

%%% FIGURE %%%%%%%%%%%%%%%%%%%%%%%%%%%%%%%%%%%%%%%%
%%%%%%%%%%%%%%%%%%%%%%%%%%%%%%%%%%%%%%%%%%%%%%%%%%%

%%% FIGURE %%%%%%%%%%%%%%%%%%%%%%%%%%%%%%%%%%%%%%%%
%%%%%%%%%%%%%%%%%%%%%%%%%%%%%%%%%%%%%%%%%%%%%%%%%%%

Next consider the behavior as the number of external gluons increases.
In \fig{fig:splitMHVs} we show relative error distributions for the
set of MHV amplitudes $A_6(1^-,2^-,3^+,4^+,5^+,6^+)$,
$A_7(1^-,2^-,3^+,4^+,5^+,6^+,7^+)$ and
$A_8(1^-,2^-,3^+,4^+,5^+,6^+,7^+,8^+)$. For each of these amplitudes
the dashed (black) curve shows the relative error in the coefficient
of the $1/\e^2$ singularity. Similarly, the relative errors in the $1/\e$
and $\ep^0$ contributions are given by the solid (red) curve and
shaded (blue) distribution. The relative precision of the $1/\e^2$
singularities is better than $10^{-11}$ for these six-, seven- 
and eight-point amplitudes. 
The computational-stability scaling properties in going from six- to
seven- and then eight-point amplitudes in fig.~\ref{fig:splitMHVs} are
also rather striking.  There is little change in the shape of the
curves as we increase the number of legs.

%%%%%%%%%%%% TABLE %%%%%%%%%%%%%%%%%%%%%%%%%%
\begin{table*}
\caption{The average time needed to evaluate one point in phase space
for various helicity configurations.  The time is in milliseconds on a
2.33 GHz Xeon processor. The second column gives the average evaluation
time for the cut part, including the recomputation at higher precision
of points identified as unstable.  The third column gives the time for
the full amplitude, including rational terms, using only ordinary
double precision.  The fourth column gives the average time 
using ordinary double precision on stable points and 
higher precision on contributions
marked as unstable either by the $1/\eps$ consistency test
(\ref{BubbleSingularity}) or the spurious-pole test (\ref{littleIR}).
\label{TimingTable} }
\vskip .4 cm
\begin{tabular}{||l|r|r|r||}
\hline
\hskip .75 cm helicity & \hskip .1 cm  cut part \hskip .1 cm  & 
                full amplitude \hskip .4 cm\null   & 
                full amplitude  \hskip .4 cm \null \\[-10pt] 
 &   &  \hskip .1 cm double prec. only  \hskip .1 cm \null& 
               \hskip .1 cm with multi-prec.  \hskip .1 cm  \\

\hline
\hline
$\; {-}{-}{+}{+}{+}{+}$ & 2.4 ms~~ & 7 ms ~~~~~~~ & 11 ms~~~~~~~ 
 \\
\hline
$\; {-}{-}{+}{+}{+}{+}{+}$ & 4.2 ms~~ &  11 ms~~~~~~~~  & 23 ms~~~~~~~ 
 \\
\hline
$\;{-}{-}{+}{+}{+}{+}{+}{+}\;$ & 6.1 ms~~ & 29 ms~~~~~~~~  & 43 ms~~~~~~~ 
 \\
\hline
$\; {-}{+}{-}{+}{+}{+}$ & 3.1 ms~~ & 18 ms~~~~~~~~  &32 ms~~~~~~~ 
 \\
\hline
$\; {-}{+}{+}{-}{+}{+}$ & 3.3 ms~~ &  61 ms~~~~~~~~ & 96 ms~~~~~~~ 
 \\
\hline
$\; {-}{-}{-}{+}{+}{+}$ & 4.4 ms~~ & 12 ms~~~~~~~~  & 22 ms~~~~~~~  
 \\
\hline
$\; {-}{-}{+}{-}{+}{+}$ & 5.9 ms~~ & 47 ms~~~~~~~~  & 64 ms~~~~~~~ 
 \\
\hline
$\; {-}{+}{-}{+}{-}{+}$ & 7.0 ms~~ & 72 ms~~~~~~~~ & 114 ms~~~~~~~ 
\\
\hline
\end{tabular}
\end{table*}

Even more striking is the modest increase in computation time.  As
mentioned earlier, the tree-like nature of on-shell methods leads us
to expect only mild scaling for a given helicity pattern, in stark
contrast with the rapid increase in required computational resources
for ordinary Feynman diagrams.  These expectations are borne out by
the values for the average computation time shown in
\Tab{TimingTable}.  The table shows the average time on a 2.33 GHz
Xeon processor for computing a color-ordered amplitude of a given
helicity configuration at a single phase-space point.  The first three
rows show the timing for the six-, seven- and eight-point MHV
amplitudes corresponding to \fig{fig:splitMHVs}.  Even for the
eight-point case we obtain an average evaluation time of less than 50
ms, including running the phase-space points marked as unstable at
higher precision.  It is also interesting to note the relatively
modest increase in computation time due to turning on higher precision
for unstable points, even in this initial implementation.  (The time
in the third column includes the evaluation of bubble coefficients
used in the spurious-pole test (\ref{littleIR}).)

Finally, consider the six-gluon NMHV amplitudes.
Figure~\ref{fig:NMHVs} illustrates the numerical stability properties
of the complete set of independent six-gluon NMHV amplitudes not
related by symmetries, $A_6(1^-,2^-,3^-,4^+,5^+,6^+)$,
$A_6(1^-,2^-,3^+,4^-,5^+,6^+)$ and $A_6(1^-,2^+,3^-,4^+,5^-,6^+)$,
compared against a quadruple-precision target computed with {\tt
BlackHat}. For each one of these amplitudes, the contributions to the
$1/\e^2$, $1/\e$ and finite $\e^0$ terms are shown in a similar format
as the MHV case.  These NMHV curves are all shifted to the right
compared to the MHV cases in in \fig{fig:splitMHVs}. This property is
not surprising; it is due to the more complicated nature of the NMHV
amplitudes.  In particular, the amplitudes contain higher powers of
the box Gram determinants in denominators of the box coefficients,
which then feed into triangle, bubble and rational contributions.  As
in the MHV cases, when one goes from $1/\ep^2$ to $1/\ep$ to $\ep^0$,
the curves shift to the right again, reflecting the more complicated
calculations.  Nevertheless, they all exhibit excellent numerical
stability, with the distributions of relative errors for the finite
pieces peaking at $10^{-8}$ or better.
We identify points as unstable, and automatically recompute such points
at higher precision, using the same criteria as for the MHV
amplitudes.  In the NMHV case, the fall-off is not as sharp as in the
MHV case.  Nevertheless, the accuracy obtained is more than sufficient
for use in an NLO program.

The average evaluation time in the current version,
for all independent six-gluon helicity configurations needed at NLO, 
including the NMHV ones, is given in \Tab{TimingTable}.
One can see that alternating-helicity configurations
do take longer to compute.  However, in all cases the cut parts are 
evaluated in under 8 ms and the full amplitudes in under 120 ms.
Although we have not
run systematic tests of NMHV amplitudes beyond six points, initial
studies at seven points indicate that the scaling behavior of the NMHV
amplitudes is not quite as good as for the MHV case, but still very good.

\section{Conclusions}
\label{ConclusionsSection}

In this paper we presented the first results from \BlackHat{}, an
automated implementation of on-shell methods, focusing on the key
practical issues of numerical stability and computational time.  We
illustrated the numerical stability by computing a variety of complete
six-, seven- and eight-gluon helicity amplitudes and comparing the
results against previously-obtained analytic results or against higher
precision calculations.  In this initial version we achieved
reasonable speed, an average computation time of 114 ms per
phase-space point
for the most complicated of the six-gluon helicity amplitudes, and
substantially better for the simpler helicities. We expect this speed
and stability to be sufficient for carrying out phenomenological
studies of backgrounds at the LHC, even as we expect further
improvements with continuing optimization of the code.  After the code
is stable and tested for a wide variety of processes, we plan to make
it publicly available.

\BlackHat{} uses the unitarity method with four-dimensional loop
momenta~\cite{UnitarityMethod,Fusing}.  This method allows the use of compact
tree-level helicity amplitudes as the basic building blocks.  We
compute the box coefficients using quadruple cuts~\cite{BCFUnitarity}.
For box integrals with massless internal propagators and at least one
massless corner, we presented a simple solution to the cut conditions.
The solution makes manifest the presence of square roots, rather than
full powers, of a spurious (Gram determinant) singularity for each
power of the loop momentum in the numerator.  We evaluated the
triangle- and bubble-integral coefficients using Forde's
approach~\cite{Forde} to expose their complex-analytic structure.
Another important ingredient in our
procedure is the OPP~\cite{OPP} subtraction of boxes from triple cuts
when computing triangle coefficients, and of boxes and triangles from
ordinary (double) cuts when computing bubble coefficients.  Viewed in
terms of Forde's complex-valued parametrization approach, the OPP
subtraction cleans the complex plane of poles, using
previously-computed coefficients.
We then introduced a discrete Fourier projection, as an efficient and
numerically stable method for extracting the desired coefficients.  
In the bubble case, this procedure can be recast in terms of spherical
harmonics.  

We computed the purely rational terms using loop-level on-shell
recursion, modifying the treatment of spurious singularities compared
to refs.~\cite{Bootstrap,Genhel}.  We used a discrete Fourier
sum to compute the spurious-pole residues from the cut
parts. These contributions are then subtracted from the
recursively-computed rational terms in order to cancel spurious singularities
implicit in the latter, and thereby make the full amplitude free of
spurious singularities as required.

The computation of most points in phase space proceeds using ordinary
double-precision arithmetic to an accuracy of $10^{-5}$ or less.
This is far better than the Monte-Carlo integration errors that will 
inevitably arise in any use of amplitudes in an NLO parton-level or 
parton-shower code (not to mention parton distribution, scale, 
shower and hadronization uncertainties).
Nonetheless, the computation of the amplitude at a small percentage of
phase-space points does manifest a loss of precision, resulting in an
instability and larger error.  In order to identify such unstable
points as may arise, we impose the requirements that all spurious
singularities cancel amongst bubble coefficients, and that the
coefficients of the $1/\e$ singularity (corresponding to $\e$-singular
terms in bubble integrals) be correct.  Whenever the calculation at a
given phase-space point fails these criteria we simply recalculate the
point at higher precision.  There are other possible means for dealing
with Gram-determinant singularities~\cite{FJT,OtherGramMethods,
GramInterpolate,Denner,NLMLesHouches}, but we prefer this approach
because of its simplicity~\cite{CutTools}.  In practice, 
it has a relatively modest impact on the overall speed of
the program.  In the most complicated of the six-gluon helicity
amplitudes, higher-precision evaluation causes the time to increase 
modestly, from 72 ms to 114 ms.
We expect to see further improvements with additional refinements.

It is important to validate a numerical method against known analytic
results.  For this purpose, we made use of MHV configurations, which
contain two gluons of helicity opposite to that of the others.  In
particular, we considered the case where the two opposite helicities
are nearest neighbors in the color order.  In earlier work, these
amplitudes were computed for an arbitrary number of external
gluons~\cite{Bootstrap,OneLoopMHV}, using on-shell methods.  We used
these results to confirm that \BlackHat{} returns the correct values
through eight gluons.  We also verified numerical stability
for non-MHV amplitudes by comparing results for
all six-gluon amplitudes against a reference computation done entirely
using quadruple-precision arithmetic.

We defer discussion of amplitudes with external fermions, or with massive
quarks and vector bosons, to the future.  (Some work directly
relevant to the question of adding massive particles may be found
in refs.~\cite{Massivetree,BFMassive,KilgoreCut}.)  We will 
also present further details, including the integral expansions we 
use around spurious singularities, in a future publication~\cite{Future}.

The excellent numerical stability and timing
performance of \BlackHat{} is due to a variety of ideas described 
in this paper.  Because the unitarity method uses
gauge-invariant tree amplitudes as the basic input into the
calculation, we avoid the large gauge cancellations inherent in
Feynman-diagram calculations.  In addition we made use of very compact
four-dimensional tree-level helicity amplitudes as the basic input to
the calculations. All steps in our computation of the rational terms,
as well as the integral coefficients, are carried out in four
dimensions.  Our simple quadruple-cut solution
(\ref{MasslessSolution}) also helps maintain numerical stability
in the box contributions. Our parametrization choices for triple and
double cuts, and the OPP subtraction of previously-computed
coefficients are additional important ingredients.  Finally, our use
of discrete Fourier projections helps considerably.

The resulting C++ code \BlackHat{} is efficient and numerically
stable, as we have illustrated with the computation of various
one-loop gluon amplitudes and their comparison to known analytic
expressions.  Based on the results presented here, we expect
\BlackHat{} to make possible the computation of a wide variety of new
one-loop amplitudes for collider physics that have been inaccessible
with traditional methods.  We hope that \BlackHat, in
conjunction with automated programs~\cite{GK} for combining the real
and virtual contributions at NLO, will soon enable the computation of
phenomenologically important cross-sections at the LHC.

\section*{Acknowledgments}

We would like to thank John Joseph Carrasco, Tanju Gleisberg,
Henrik Johansson and Michael Peskin for helpful discussions.
We are grateful to the
Galileo Galilei Institute for Theoretical Physics for hospitality
while part of this work was carried out.  We also
thank Academic Technology Services at UCLA for computer support.  This
research was supported by the US Department of Energy under contracts
DE--FG03--91ER40662 and DE--AC02--76SF00515.  CFB's research was
supported in part by funds provided by the U.S. Department of Energy
(D.O.E.) under cooperative research agreement DE-FC02-94ER40818, and
in part by the National Science Foundation under Grant
No. PHY05-51164.  DAK's research is supported by the Agence Nationale
de la Recherce of France under grant ANR--05--BLAN--0073--01.
The work of DM was supported by the Swiss National Science Foundation
(SNF) under contract PBZH2-117028.

%%%%%%%%%%%%%%%%%%%%%%%%%%%%%%%%%%%%%%%%%%%%%%%%%%%%%%%%%%%%%%%%%%%%%

\end{document}